\documentclass[acmsmall]{acmart}

\usepackage{float}
\usepackage{subfig}
\usepackage{multirow}
\usepackage{graphicx}
\usepackage{enumitem}
\usepackage{colortbl}
\usepackage{array}
\usepackage{soul}
\usepackage{amsmath}
\usepackage{color}

\AtBeginDocument{%
  \providecommand\BibTeX{{%
    \normalfont B\kern-0.5em{\scshape i\kern-0.25em b}\kern-0.8em\TeX}}}
\usepackage{amsmath}
\usepackage{bm}

\setcopyright{acmcopyright}
\copyrightyear{2024}
\acmYear{2024}
\acmDOI{XXXXXXX.XXXXXXX}

\begin{document}

\settopmatter{printacmref=false} 

\newcommand{\fullname}{\emph{\textbf{Group-\underline{A}lignment and Global-\underline{U}niformity Enhanced \underline{R}epresentation \underline{L}earning for Debiasing Recommendation~(AURL)}}}
\newcommand{\shortname}{\emph{AURL}}

\title[Group-Alignment and Global-Uniformity for debiasing RS]{Mitigating Recommendation Biases via Group-Alignment and Global-Uniformity in Representation Learning}

\author{Miaomiao Cai}
\affiliation{%
  \institution{Hefei University of Technology}
  \city{Hefei}
  \country{China}
}
\email{cmm.hfut@gmail.com}

\author{Min Hou}
\affiliation{%
  \institution{Hefei University of Technology}
  \city{Hefei}
  \country{China}
}
\email{hmhoumin@gmail.com}

\author{Lei Chen}
\affiliation{%
  \institution{Tsinghua University}
  \city{Beijing}
  \country{China}
}
\email{chenlei.hfut@gmail.com}

\author{Le Wu}
\authornote{Corresponding Authors.}
\affiliation{%
  \institution{Hefei University of Technology}
  \city{Hefei}
  \country{China}
}
\email{lewu.ustc@gmail.com}

\author{Haoyue Bai}
\affiliation{%
  \institution{Hefei University of Technology}
  \city{Hefei}
  \country{China}
}
\email{baihaoyue621@gmail.com}

\author{Yong Li}
\affiliation{%
  \institution{Tsinghua University}
  \city{Beijing}
  \country{China}
}
\email{liyong07@tsinghua.edu.cn}

\author{Meng Wang}
\authornotemark[1]
\affiliation{%
  \institution{Hefei University of Technology}
  \city{Hefei}
  \country{China}
}
\email{eric.mengwang@gmail.com}

\setcopyright{acmlicensed}
\acmJournal{TIST}
\acmYear{2024} \acmVolume{1} \acmNumber{1} \acmArticle{1} \acmMonth{1}\acmDOI{10.1145/3664931}

\renewcommand{\shortauthors}{Group-Alignment and Global-Uniformity Enhanced Representation Learning for Debiasing Recommendation}

\begin{abstract}
Collaborative Filtering~(CF) plays a crucial role in modern recommender systems, leveraging historical user-item interactions to provide personalized suggestions. However, CF-based methods often encounter biases due to imbalances in training data. This phenomenon makes CF-based methods tend to prioritize recommending popular items and performing unsatisfactorily on inactive users.
Existing works address this issue by rebalancing training samples, reranking recommendation results, or making the modeling process robust to the bias.
Despite their effectiveness, these approaches can compromise accuracy or be sensitive to weighting strategies, making them challenging to train.
Therefore, exploring how to mitigate these biases remains in urgent demand.

In this paper, we deeply analyze the causes and effects of the biases and propose a framework to alleviate biases in recommendation from the perspective of representation distribution, namely ~ \fullname.
Specifically, we identify two significant problems in the representation distribution of users and items, namely group-discrepancy and global-collapse.
These two problems directly lead to biases in the recommendation results.
To this end, we propose two simple but effective regularizers in the representation space, respectively named group-alignment and global-uniformity.
The goal of group-alignment is to bring the representation distribution of long-tail entities closer to that of popular entities, while global-uniformity aims to preserve the information of entities as much as possible by evenly distributing representations.
Our method directly optimizes both the group-alignment and global-uniformity regularization terms to mitigate recommendation biases.
Please note that ~\shortname ~ applies to arbitrary CF-based recommendation backbones. 
Extensive experiments on three real datasets and various recommendation backbones verify the superiority of our proposed framework. The results show that ~\shortname ~ not only outperforms existing debiasing models in mitigating biases but also improves recommendation performance to some extent.
\end{abstract}



\maketitle

\section{INTRODUCTION}
Personalized recommendations have become indispensable in various online applications, serving as valuable tools for users to cope with information overload~\cite{ chen2019efficient,yang2023generative,hou2019explainable}.
As the most popular schema for personalized recommender systems, Collaborative Filtering~(CF)~\cite{Rendle2009BPRBP, He2020LightGCNSA, Yu2021AreGA} utilizes similarities between users and items hidden in historical user-item interactions to provide recommendations. 
Typically, CF-based methods encode users and items into a shared space and then recover the user-item interactions (preferences) through the corresponding representations~\cite{Wei2020ModelAgnosticCR}.

\begin{figure}[t]
    \vspace{-0.5cm}
    \centering
    \subfloat[Item popularity bias]{\includegraphics[width = 0.4\linewidth]{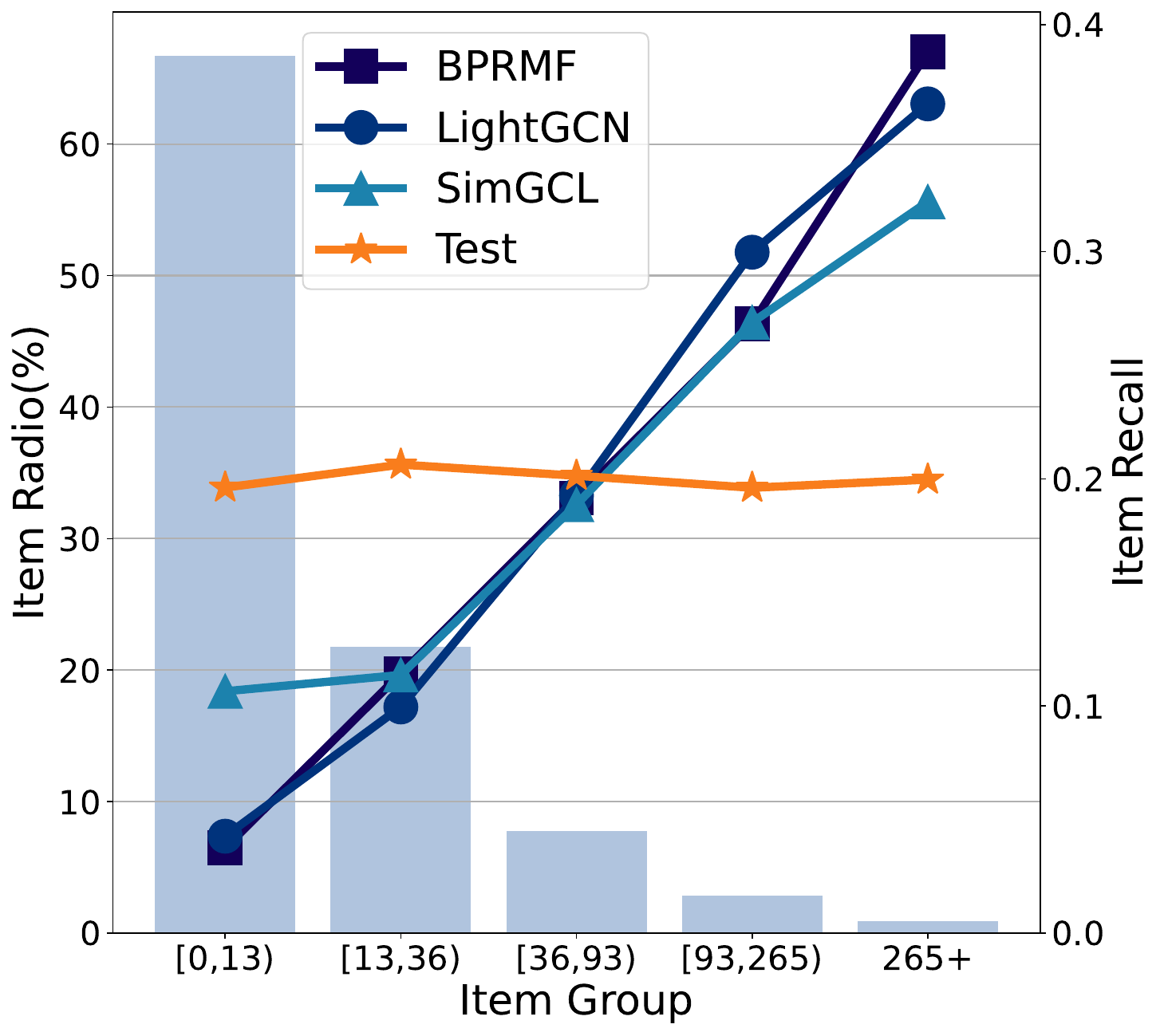}\label{Item popularity bias}}\hspace{10mm}
    \subfloat[User consistency bias]{\includegraphics[width = 0.4\linewidth]{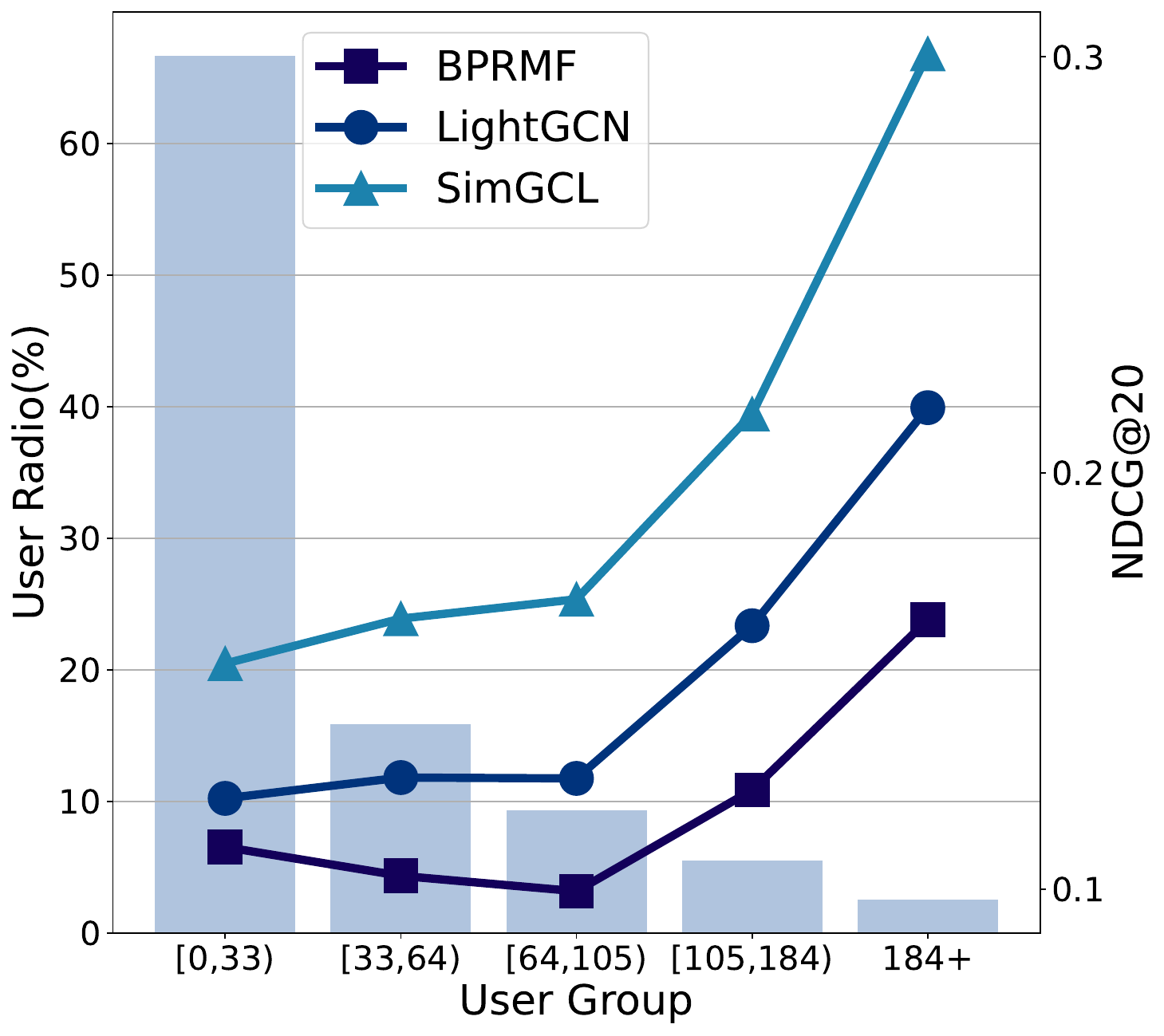}\label{User consistency bias}}
    \vspace{-0.3cm}
    \caption{We analyze biases in the results of three typical CF models—BPRMF~\cite{Rendle2009BPRBP}, LightGCN~\cite{He2020LightGCNSA}, and SimGCL~\cite{Yu2021AreGA} on the Douban-Book dataset~\cite{Yao2020SelfsupervisedLF}. To facilitate our illustration, we categorize items and users into groups based on their popularity in the training set. We then evaluate, based on the TopK recommendation lists, the recommendation frequency ($Item ~ Recall$) for each item group and the accuracy performance ($NDCG@20$) for each user group.}
    \Description{Item popularity bias.}
    \label{fig: bias in results}
    
    \vspace{-0.5cm}
\end{figure}

Although CF-based methods have achieved considerable success, their training approach, which involves reconstructing historical interactions, makes them susceptible to imbalances within the training data, leading to biases in recommendation results~\cite{Wei2020ModelAgnosticCR, Chen2020BiasAD, Zhu2021PopularityOpportunityBI}.
In real-world scenarios, the frequency distribution of users and items in training interactions is uneven. They often follow a power-law distribution, where a small number of popular items and active users dominate the majority of interactions~\cite{Wei2020ModelAgnosticCR}.
This phenomenon causes CF-based methods to prioritize recommending popular items and perform unsatisfactorily on inactive users, thereby failing to uncover the real characteristics of items and the true preferences of users~\cite{Chen2020BiasAD}.
Fig. \ref{fig: bias in results} shows the item popularity bias and user consistency bias on the real-world Douban-Book dataset~\cite{Yao2020SelfsupervisedLF}. 
We group items and users according to their frequency (popularity) of appearance in the training data, and the background histograms indicate the ratio of items/users in each group~\cite{Wei2020ModelAgnosticCR}. We can observe a clear phenomenon of data imbalance. Then we train the mainstream CF-based models BPRMF~\cite{Rendle2009BPRBP}, LightGCN~\cite{He2020LightGCNSA} and SimGCL~\cite{Yu2021AreGA}. We count the frequency of items in the recommendation results for each item group (Fig. \ref{fig: bias in results}(a)) and calculate the recommendation performance for each user group (Fig. \ref{fig: bias in results}(b)). In Fig. \ref{fig: bias in results}(a), the orange line shows the real item frequency in the test dataset. As evident, items that are more popular in the training data are recommended far more frequently than anticipated, highlighting a significant item popularity bias. 
In Fig. \ref{fig: bias in results}(b), there is inconsistency in the effectiveness of the recommendations between different user groups, and inactive users experience unsatisfactory recommendations.
These biases significantly impact the performance of recommender systems, undermining both the diversity of recommendations and the user experience~\cite{Chen2020BiasAD, AutoDebias, Wei2020ModelAgnosticCR}.
Even more concerning, item popularity bias can cause the "Matthew effect", where popular items receive more recommendations and consequently become even more popular~\cite{Wei2020ModelAgnosticCR}.

Given the significant impact of biases, debiasing in recommender systems has recently become a key area of research~\cite{AutoDebias, Chen2020BiasAD, Zhu2021PopularityOpportunityBI, Zhao2022InvestigatingAP,fairdata23,zhao2023fair,gao2023cirs}. Previous efforts have addressed bias mitigation from several angles:
(1) Sampling Strategies: Some studies down-sample~\cite{fairdata23} popular entities (items and users) or up-sample~\cite{Zhu2021PopularityOpportunityBI} unpopular ones to balance interaction distributions in the training set.
(2) Robust Methods: Techniques such as causal inference~\cite{Wei2020ModelAgnosticCR,zheng2021disentangling} and adversarial training~\cite{fairgo,AutoDebias} have been employed to enhance model robustness against biased data~\cite{Zhu2021PopularityOpportunityBI, Zhao2022InvestigatingAP}.
(3) Post-Processing: Certain approaches re-rank recommendation lists to prevent the over-recommendation of popular items\cite{Zhu2021PopularityOpportunityBI,Chen2020BiasAD}.
While these methods are commendable, they exhibit imperfections, such as potentially compromising recommendation accuracy by oversimplifying data balancing~\cite{Chen2020BiasAD}, which can ignore users' true preferences. Causal inference methods\cite{Wei2020ModelAgnosticCR}, requiring stringent data generation assumptions, and adversarial training~\cite{Zhu2021PopularityOpportunityBI}, which introduces instability by blurring distinctions between popular and unpopular entities, illustrate the challenges of existing approaches~\cite{zheng2021disentangling,fairgo}. Therefore, further exploration into effective bias mitigation remains critically necessary.

\begin{figure}[t]
    \vspace{-0.5cm}
    \centering
    \subfloat[Item representation distribution]{\includegraphics[width = 0.5\linewidth, height=0.3\textwidth]{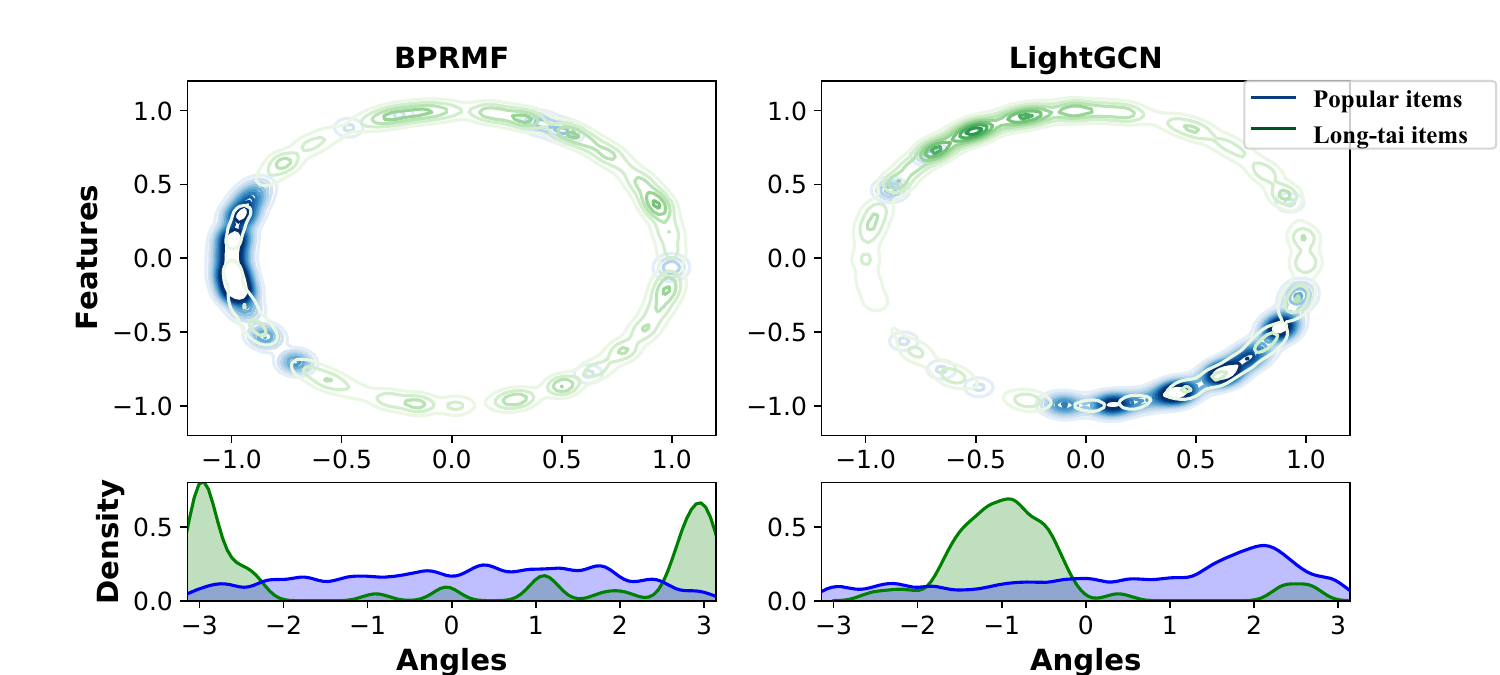}\label{Item representation distribution}}
    \subfloat[User representation distribution]{\includegraphics[width = 0.5\linewidth, height=0.3\textwidth]{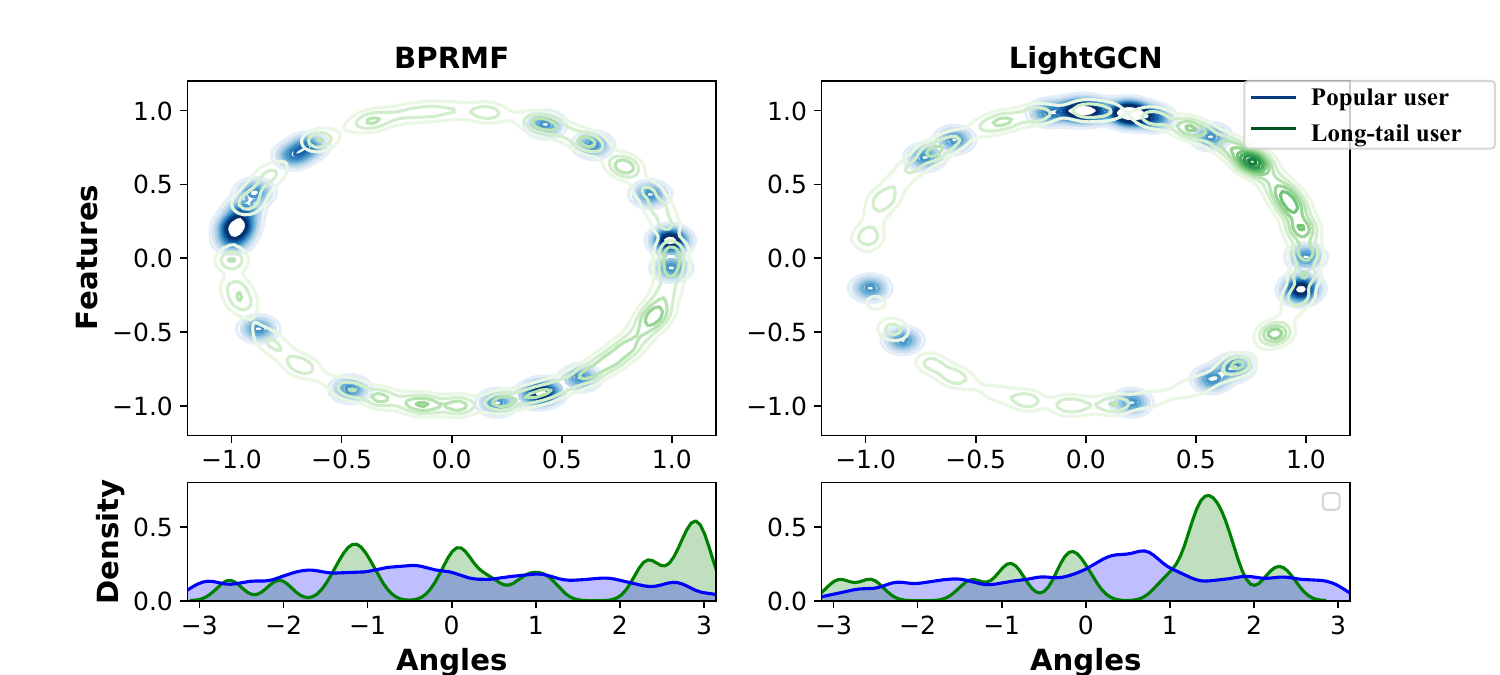}\label{User representation distribution}}
    \vspace{-0.3cm}
    \caption{Representation distribution of the Douban-Book dataset on $\mathcal{S}^1$. We plot the representation distributions using Gaussian Kernel Density Estimation (KDE) in $\mathbb{R}^2$ and von Mises-Fisher (vMF) KDE on angles (i.e., $\arctan2(y, x)$ for each point $(x, y)$ on $\mathcal{S}^1$). Specifically, we categorize items and users into two groups based on their popularity: blue represents popular items/users, while green denotes unpopular items/users.}
    \label{fig:representation_distribution}
    \Description{Representation distribution of Douban-Book on $\mathcal{S}^1$.}
    \vspace{-0.5cm}
\end{figure}

In fact, it is known that the quality of learned representations plays a crucial role in the recommendation.
Most CF-based recommendation models can be divided into two parts~\cite{citationsurveylekey}. 
Firstly, encoders project users and items into a representation space~\cite{Koren2009MatrixFT}. Then, an interaction function computes preferences between users and items in the space~\cite{WWW2017NCF}. The interaction function is usually set as the simple inner product, while researchers meticulously design various kinds of encoders to make the representations as informative as possible~\cite{Wang2022TowardsRA}.
Consequently, we highlight that \textbf{addressing the impact of biases from the perspective of representation distribution and obtaining better representations} contains a huge potential to effectively resolve the issue of biases.
However, achieving this goal is not trivial.  Initially, scrutinizing the impact of biases on representations is imperative.
In Fig.\ref{fig:representation_distribution}, we respectively map the learned representations of two typical CF-based models as 2-dimensional normalized vectors on the unit hypersphere $\mathcal{S}^1$ (i.e., the circle with radius 1). Then, we plot the feature distribution using nonparametric Gaussian Kernel Density Estimation (KDE) in $\mathbb{R}^2$ and visualize the density estimations at angles for each entity on $\mathcal{S}^1$.
According to Fig.\ref{fig:representation_distribution}, we can observe notably different feature/density distributions between popular entities and long-tail entities.
We believe that this phenomenon arises from the imbalanced data, which results in a scarcity of interactions for long-tail entities, thereby hindering the acquisition of accurate representations. 
Consequently, the distributions of representations for long-tail entities exhibit inconsistency when compared to those of popular entities with adequate interactions. Ideally, there should be no distributional shift between the popular entities and long-tail entities. We simplify this phenomenon as \textbf{group-discrepancy}.
Furthermore, we find that the distribution of entity representations exhibits a folding pattern, clustering around several points.
This clustering phenomenon indicates that the learned representations lack informativeness and fail to capture the distinctive characteristics of each entity~\cite{wang2020understanding}.
We name the drawback as \textbf{global-collapse}. Addressing the two impacts of bias on representation distribution is the key to mitigating recommendation biases.

In this paper, we advocate for a paradigm shift in analyzing and addressing both item-side popularity bias and user-side consistency bias from the perspective of representation distribution.
We begin by investigating the causes of group-discrepancy and global-collapse through both mathematical and empirical analyses.
To address these two issues, we propose two simple but effective regularizers in representation space respectively, named group-alignment and global-uniformity.
Specifically, the group-alignment regularizer aims to bring the representation distribution of long-tail entities closer to that of popular entities.
This regularizer transfers knowledge from the well-trained representations of popular entities to those of long-tail entities, thereby enhancing the latter's representation quality.
Inspired by the uniformity property in contrastive learning~\cite{wang2020understanding}, we design a uniformity-based regularize. 
The regularizer leads the representations roughly uniformly distributed on the unit hypersphere, preserving as much information of the entities as possible. 
To this end, we propose a framework from the representation distribution, namely ~\fullname.
Our methodology directly optimizes the group-alignment and global uniformity regularization terms to reduce group-discrepancy and global-collapse, thereby further mitigating bias issues in recommendation results. Extensive testing on three real-world datasets confirms the efficacy of our method in reducing biases and enhancing recommendation accuracy across various CF-based models.
The main contributions of this paper are summarized as follows:
\begin{enumerate}
    \item We advocate a novel perspective, utilizing representation distribution, to address both item-side popularity bias and user-side consistency bias, supported by comprehensive mathematical analyses and empirical evidence.
    \item We design the group-alignment and global-uniformity regularizers to effectively counter the biases induced by group-discrepancy and global-collapse, respectively.
    \item Extensive testing on three real-world datasets verifies the efficacy of our method in reducing biases and improving recommendation accuracy across various CF-based models.
\end{enumerate}

\section{Related work}
\subsection{Collaborative filtering}
Collaborative Filtering (CF) is a widely used technique in recommender systems, aiming to provide personalized recommendations by leveraging users' preferences and behaviors~\cite{Rendle2009BPRBP}.
The basic idea behind CF is that users with similar interests and preferences are likely to have similar opinions about items~\cite{Koren2009MatrixFT}. 
MMost CF-based models can be divided into two components~\cite{citationsurveylekey}. Firstly, encoders project users and items into a representation space~\cite{Koren2009MatrixFT}. Subsequently, an interaction function computes the preferences between users and items within this space. The interaction function is typically set as a simple inner product, while researchers meticulously design various encoders to make the representations as informative as possible~\cite{Wang2022TowardsRA}.

The simplest encoders can directly map user and item IDs into the representation space~\cite{Koren2009MatrixFT, Rendle2009BPRBP}. With the development of deep learning, neural-based encoders such as multi-layer perceptrons~\cite{WWW2017NCF} and attention mechanisms~\cite{attchen2017attentive} have emerged in recent years to capture the complex relationships between users and items.
User-item interaction data can naturally be organized into a bipartite graph, prompting researchers to employ Graph Neural Networks~(GNNs) ~\cite{Wang2019NeuralGC,deng2022graph,chen2024graph,wu2022graph} to encode more accurate node representations and high-order structural information. For example, NGCF~\cite{Wang2019NeuralGC}, LR-GCCF~\cite{Chen2020RevisitingGB}, and LightGCN~\cite{He2020LightGCNSA} utilize high-order relationships on interaction graphs to enhance representation performance. 
Recently, Self-Supervised Learning (SSL) has been introduced to improve the generalizability of representations~\cite{Wu2020SelfsupervisedGL}.
For example, SimGCL~\cite{Yu2021AreGA} employs noise feature enhancement methods and constructs comparison targets to enhance the accuracy and robustness of representations.
Despite their effectiveness, CF-based models often overlook biases in recommendation results due to the imbalanced distribution of interaction data~\cite{Chen2020BiasAD}.

\subsection{Debiasing methods in recommendation results}

Mitigating biases in recommendation results is a common task in recommender systems and has been extensively studied. 
Previous research work has tried to alleviate recommendation biases from multiple perspectives.
We classify the methods into re-weighting-based, decorrelation-based, and adversarial-based approaches.

\textbf{Re-weighting-based methods} aim to shift attention away from popular items/users either during training or prediction, thereby increasing the importance of unpopular items/users in the recommendation process~\cite{Zhao2022InvestigatingAP, Zhu2021PopularityOpportunityBI,AutoDebias,naghiaei2022cpfair,li2021user,cui2012discover}. 
For example, Inverse Propensity Scoring (IPS)~\cite{Zhu2021PopularityOpportunityBI} compensates for unpopular items/users by adjusting predictions within the user-item preference matrix, thereby elevating preference scores and rankings for unpopular items/users.
Explanation: Expanded the acronym "IPS" for clarity and refined the sentence structure to improve readability.
Similarly, $\gamma$-AdjNorm~\cite{Zhao2022InvestigatingAP} enhances the focus on unpopular items/users by controlling the normalization strength during the neighborhood aggregation process in GCNs-based models.
DORL~\cite{gao2023alleviating} addresses the Matthew effect in offline reinforcement learning recommendations by introducing a penalty term, mitigating the conservatism inherent in existing methods.
Zerosum~\cite{rhee2022countering} reduces model bias in recommendation systems by directly equalizing recommendation scores across items preferred by a user.

\textbf{Decorrelation-based methods} aim to mitigate the influence of popularity on item/user representations or prediction scores by removing correlations between them~\cite{Wei2020ModelAgnosticCR,fairgo,bonner2018causal,zhang2021causal,wang2022causal,ren2022mitigating}. For example, MACR~\cite{Wei2020ModelAgnosticCR} utilizes counterfactual reasoning to eliminate the direct impact of popularity on item/user outcomes. 
TIDE~\cite{evil} leverages temporal information to differentiate between benign bias due to item quality and harmful bias resulting from conformity.

\textbf{Adversarial-based methods} aim to engage in a minimax game between the recommender G and an introduced adversary D, such that D provides signals to increase the recommendation opportunities for unpopular items or enhance user accuracy~\cite{fairgo,liu2023mitigating,kusner2017counterfactual,wang2023survey,shao2022faircf,Popularity-awareDRO}.
FairGo~\cite{fairgo} enhances the graph network recommendation model by incorporating discriminators, which predict fairness-related attributes of nodes by utilizing their embeddings and the embeddings of the surrounding network structure.
FairMI~\cite{zhao2023fair} employs adversarial principles to minimize mutual information between embeddings and sensitive attributes while maximizing it between embeddings and non-sensitive information.
In contrast, InvCF~\cite{zhang2023invariant} exploits the notion that item/user representations remain unchanged despite variations in popularity semantics. By filtering out unstable or outdated popularity characteristics, InvCF learns unbiased representations.

Although previous methods have worked hard to reduce biases in recommendation results, they still have certain limitations. For example, sampling strategies and post-processing methods usually only focus on long-tail items/users, at the expense of accuracy for popular items/users~\cite{Zhu2021PopularityOpportunityBI}. Methods based on causal reasoning typically rely on strong assumptions about data generation~\cite{Wei2020ModelAgnosticCR}, and adversarial learning methods are often unstable~\cite{fairgo}. Therefore, there is still an urgent demand to explore more effective methods to address these biases in recommender systems~\cite{Chen2020BiasAD}.

\subsection{Representation learning}

Representation learning plays a crucial role in collaborative filtering by generating personalized representations for each user and item~\cite{yang2023hyperbolic}.
These personalized representations more accurately reflect the interests of users and the characteristics of items, leading to enhanced accuracy and effectiveness in CF-based models~\cite{He2020LightGCNSA, Koren2009MatrixFT,ren2022semi}.
Without effective representation learning, models may struggle to capture the underlying relationships between users and items accurately, resulting in less accurate recommendations~\cite{Wu2020SelfsupervisedGL}.

In the field of representation learning, researchers typically focus on two fundamental properties to measure the quality of learned representations: alignment and uniformity~\cite{wang2020understanding}.
The goal of alignment is to ensure that the learned representations effectively capture the similarities between positive data points~\cite{wang2020understanding}. 
This enhances the model's ability to express relationships by bringing similar data points closer together in the representation space.
In Natural Language Processing (NLP), alignment is used to align words or phrases between different languages, facilitating cross-lingual text translation and information retrieval~\cite{sebastian2023malayalam}.
In Computer Vision (CV), alignment is applied to synchronize features across images from different domains, such as aligning images of varied styles or sensors into a unified representation space for applications like image generation~\cite{zhu2021progressive}, style transfer~\cite{shen2017style}, and more.
In recommender systems~(RS), alignment can be used to align user and item representations to better capture user interests and item characteristics~\cite{Wang2022TowardsRA}.
Unlike previous studies, this research focuses on the relationship between popular entity groups and long-tail entity groups. 
By aligning the representation distributions of different groups, we aim to achieve improved group-alignment, ensuring that the distribution of entities in the representation space remains consistent, regardless of changes in popularity.

The goal of uniformity is to ensure that learned representations are evenly distributed in the feature space, typically by having these representations roughly evenly spread across the unit hypersphere~\cite{wang2020understanding,zhang2023rethinking}.
This enhances the model's generalization performance, leading to a more balanced and consistent distribution of representations~\cite{Yu2021AreGA, Wu2020SelfsupervisedGL}. 
In NLP, uniformity ensures that word representations are evenly distributed within the semantic space, improving the quality of word vectors and thereby enhancing performance in text-related tasks~\cite{zhang2023rethinking}.
In CV, uniformity ensures that feature representations are evenly distributed across the image space, helping the model more effectively capture diverse image features~\cite{zhang2023rethinking}.
In RS, uniformity ensures that the distribution of user and item representations is more informative, thereby improving both the accuracy and generalization of recommendations~\cite{Wu2020SelfsupervisedGL, Wang2022TowardsRA, Yu2021AreGA}.

Previous studies have demonstrated that InfoNCE optimizes both the alignment and uniformity of representations by aligning positive samples and distancing negative samples~\cite{Yu2021AreGA,wang2020understanding}.
However, these studies typically focus on aligning different views of the same user or item post data augmentation, which can introduce selection bias in terms of uniformity optimization.
In contrast, methods such as DirectAU~\cite{Wang2022TowardsRA} directly optimize both alignment and uniformity, thus circumventing issues related to selection bias in positive and negative sample selection. 
In this paper, we explore the issue of inconsistent distributions across different item sets, emphasizing the need to align from the perspective of feature distributions.
By integrating both alignment and uniformity, our approach effectively optimizes feature representations, enhances recommendation performance, and adapts better to variations in data distributions, thus delivering improved outcomes for recommendation systems.

\section{Impact of Bias on Representation Learning in CF}

\subsection{Debiasing problem formulation}

In this subsection, we first establish a formal definition of the debiasing problem in CF.
Let $\boldsymbol{U}$ (where $|\boldsymbol{U}| = M$) and $\boldsymbol{I}$ (where $|\boldsymbol{I}| = N$) denote the sets of users and items in the CF-based models, respectively.
Assuming the implicit feedback setting, let $\mathcal{R} \in {0,1}^{M \times N}$ represent the observed implicit interaction matrix, where $\mathcal{R}_{u,i} = 1$ if user $u$ has interacted with item $i$; otherwise, it is 0.
The key to CF-based models is accurately learning the user representation matrix $\mathbf{Z} \in \mathbb{R}^{M \times D}$ and the item representation matrix $\mathbf{H} \in \mathbb{R}^{N \times D}$, where $D$ denotes the representation size and $D \ll M, N$.
With the representation matrix, the predicted score is defined as the similarity between the user and item representation. 
Specifically, the similarity between users and items is calculated using the inner product of their representations, i.e., $s(u,i) = \mathbf{z}_u^T \mathbf{h}_i$.
Here $s(u,i)$ is the prediction score of user $u$ to the item $i$, $\mathbf{z}_u$ and $\mathbf{h}_i$ denote the representation of user $u$ and item $i$.
To directly capture information from interactions, most studies employ the Bayesian Personalized Ranking (BPR) loss~\cite{Rendle2009BPRBP}, a meticulously designed ranking objective function for recommendations. Formally, the BPR loss is as follows:
\begin{equation}
    \mathcal{L}_{BPR}= -\frac{1}{|\mathcal{R}|}\sum_{(u,i)\in \mathcal{R}, 
 i^- \in \boldsymbol{I}/\boldsymbol{I}^+_u} ln\sigma(s(u,i)-s(u,i^-)),
    \label{bprloss}
\end{equation}
where $\sigma(\cdot)$ is the sigmoid function, $\boldsymbol{I}^+_u$ represents the set of items interacted by users in the training, $i$ is the positive item that the user has interacted with, and $i^-$ is a randomly sampled negative item that the user has not interacted with.
Specifically, the BPR loss ensures that the predicted score of observed interactions is higher than that of sampled unobserved interactions. 

In this paper, focusing on the CF-based recommendation task, we investigate biases in recommendation results, specifically addressing user-side consistency bias and item-side popularity bias. 
To better define and analyze biases in recommendation results, we divide items and users into two groups based on their popularity levels, namely:
\begin{equation}
    \boldsymbol{U}=G_{pop}^{user}\cup G_{tail}^{user},
    \boldsymbol{I}=G_{pop}^{item}\cup G_{tail}^{item},
\end{equation}
where $G_{pop}^{user}$ and $G_{pop}^{item}$ represent the popular user and item groups, respectively, while $G_{tail}^{user}$ and $G_{tail}^{item}$ denote the long-tail user and item groups. All groups are mutually exclusive. In our work, we aim to ensure that all groups have fair performance.
Specifically, for user consistency bias, we expect the model to provide similar recommendation quality across user groups, regardless of their popularity levels.
We define the debiasing objective on the user side by Demographic Parity~(DP) as provided in~\cite{Chen2020BiasAD} ~:
\begin{equation}    
    \mathbb{E}_{u \in G_{pop}^{user}}[ACC(u)]=\mathbb{E}_{u \in G_{tail}^{user}}[ACC(u)],
\end{equation}
where $ACC(u)$ define the recommendation accuracy of user $u$.
In contrast to user-side bias, for item popularity bias, we aim for each item group to have equal opportunities for exposure in the TopK recommendations, formally stated as:
\begin{equation}
    p(G_{pop}^{item}|TopK)=p(G_{tail}^{item}|TopK),
\end{equation}
where $p(G_{pop}^{item}|TopK)$ and $p(G_{tail}^{item}|TopK)$ respectively represent the probability of popular item and long-tail item being in the TopK recommendation list.

In summary, the primary objective of this paper is to mitigate biases in recommendation results, aiming to ensure fairness across different entity groups and reduce disparities in those results.
Additionally, we aim to preserve the competitive advantage of the recommendation model while minimizing any adverse effects on its overall effectiveness.

\begin{figure}[t]
    \centering
    \vspace{-0.5cm}
    \includegraphics[width=0.9\linewidth]{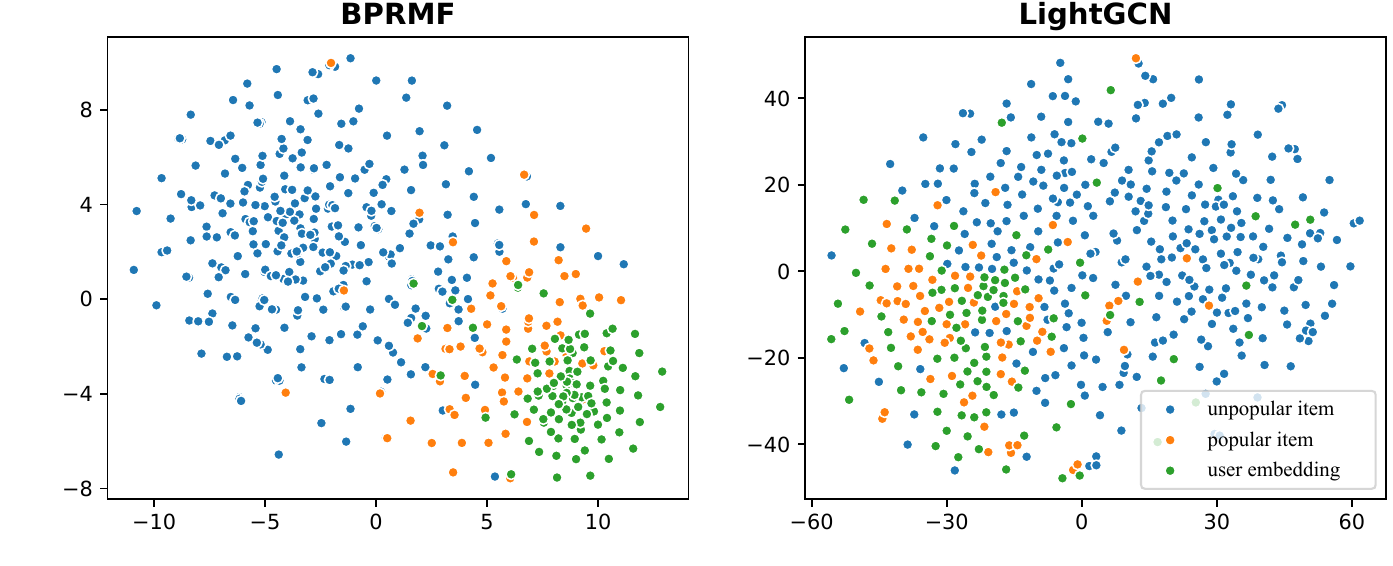}
    \vspace{-0.3cm}
    \caption{Representation visualization of items and users in the Douban-Book dataset. We randomly selected 500 items and 200 users and utilized T-SNE to visualize the representation spaces of BPRMF and LightGCN, respectively. In the visualization, green dots represent users, while blue and orange dots represent popular and long-tail items, respectively. It is evident that the number of long-tail items significantly exceeds that of popular items.}
    \Description{Representation visualization of items and users in Douban-Book.}
    \label{fig:fc}
    \vspace{-0.5cm}
\end{figure}

\subsection{Biases in representation distribution}
In this subsection, we conduct an in-depth analysis of the causes and manifestations of the biases in representation distribution. 
We take the most popular BPR loss as an example to analyze the impact of biases.
Formally, for a training sample $(u,i) \in \mathcal{R}$, the BPR loss is defined as:
\begin{equation}
    \mathcal{L}_{(u,i,i^-)}=-ln\sigma(s(u,i)-s(u,i^-))=-ln\sigma(\mathbf{z}_u^T\mathbf{h}_i-\mathbf{z}_u^T \mathbf{h}_{i^-}),~where~ i^- \in \boldsymbol{I}/\boldsymbol{I}_u^+.
\end{equation}
To optimize the BPR loss function using Stochastic Gradient Descent (SGD)~\cite{amari1993backpropagation}, we calculate the gradients of the user representation $\mathbf{z}u$, positive sample representation $\mathbf{h}i$, and negative sample representation $\mathbf{h}{i^-}$ as follows:

\begin{equation}
    \nabla_{z_{u}}=\frac{\partial \mathcal{L}_{(u,i,i^-)}}{\partial \mathbf{z}_u}=-(1-\sigma(s(u,i)-s(u,i^-)))(\mathbf{h}_i-\mathbf{h}_{i^-}),
    \label{user gradient}
\end{equation}

\begin{equation}
    \nabla_{\mathbf{h}_{i}}=\frac{\partial \mathcal{L}_{(u,i,i^-)}}{\partial \mathbf{h}_i}=-(1-\sigma(s(u,i)-s(u,i^-)))\mathbf{z}_u,
    \label{positive gradient}
\end{equation}

\begin{equation}
    \nabla_{\mathbf{h}_{i^-}}=\frac{\partial \mathcal{L}_{(u,i,i^-)}}{\partial \mathbf{h}_{i^-}}=(1-\sigma(s(u,i)-s(u,i^-)))\mathbf{z}_u.
    \label{negative gradient}
\end{equation}
According to these equations, it is evident that the gradient update directions for the positive sample representation $\mathbf{h}i$ and the negative sample representation $\mathbf{h}{i^-}$ are diametrically opposite. As recommendation data typically follows the power-law distribution, positive items are often popular items sampled from $(u,i) \sim \mathcal{R}$. Conversely, negative samples are usually drawn randomly from the entire itemset $\boldsymbol{I}$, typically resulting in long-tail items. Consequently, when using SGD to optimize the BPR loss, popular items and long-tail items tend to be updated to distinct positions within the representation space:
\begin{equation}
    \mathbf{h}_{i} \gets \mathbf{h}_{i} - \eta \nabla_{\mathbf{h}_{i}}, ~ ~\mathbf{h}_{i^-} \gets \mathbf{h}_{i^-} - \eta \nabla_{\mathbf{h}_{i^-}},
\end{equation}
where $\eta$ is the learning rate, and typically $i \in G{pop}^{item}$ and $i^- \in G_{tail}^{item}$. As a result, the distribution of item groups in the representation space becomes inconsistent. Furthermore, it is observed that the update direction for users aligns with that of the positive sample, leading most users to cluster near popular items while distancing from long-tail items.

To intuitively understand the impact of biases on representation distribution, we visualize the representations of users and items using two common CF-based models, BPRMF~\cite{Rendle2009BPRBP} and LightGCN~\cite{He2020LightGCNSA}.
We map the learned representations into a two-dimensional (2D) space using t-SNE~\cite{van2008visualizing}, ensuring all representations are captured at the point of optimal performance.
From Fig.~\ref{fig:fc}, we observe a distinctly different distribution of users and items.
Consistent with our analysis, popular items and long-tail items are located in separate regions of the representation space, with user representations predominantly clustering around popular items.
The phenomenon of inconsistent distribution may be the reason why the results of different groups are slightly different.
This distribution pattern biases the model towards recommending popular items over considering users' actual preferences and the attributes of the items.

Furthermore, the discussion above highlights that the optimization directions for different entities are all aligned with the representations of positive entities.
The frequent appearance of popular entities in the training set homogenizes the optimization direction of the representations, meaning they are optimized in similar directions.
As illustrated in Fig.~\ref{fig:representation_distribution}, representations from different groups tend to cluster around several focal points.  
This clustering indicates that the differences between representations are minimal, rendering the representations less informative.
Furthermore, this clustering causes the representations of different groups to diverge markedly, further exacerbating the bias towards popular entities.
Based on the foregoing discussion, we summarize the impact of bias on representations as follows:

\begin{itemize}[itemsep=2pt,topsep=0pt,parsep=2pt]

    \item [$\bullet$]\textbf{Group-Discrepancy}: The representations of various groups' entities are localized to distinct regions within the representation space, indicating a segregation based on group characteristics.
    
    \item [$\bullet$]\textbf{Global-Collapse}: The distribution of user and item representations shows a folding pattern, with the data densely clustering around a few focal points, leading to reduced representational diversity and potential loss of information.
        
\end{itemize}

\section{The proposed model}
\begin{figure}[t]
    \centering
    \vspace{-0.5cm}
    \includegraphics[width=0.99\linewidth]{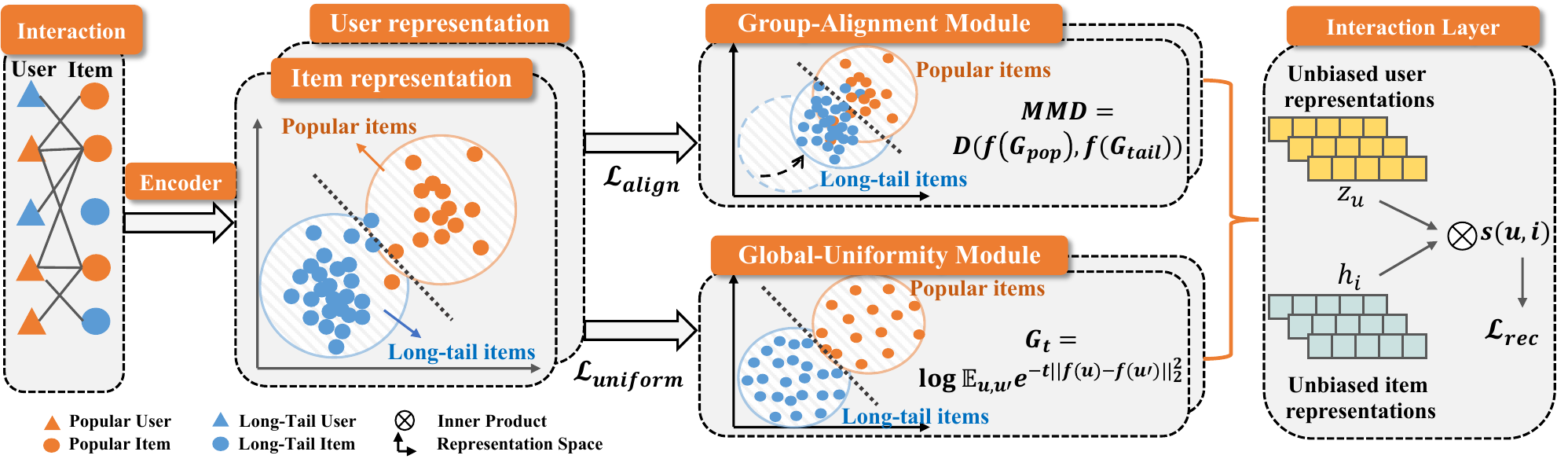}
    \vspace{-0.2cm}
    \caption{An illustration of the ~\shortname~ framework begins with input data being encoded through CF-based models to map users and items into the representation space. Subsequently, these representations are constrained by two modules: group-alignment $\mathcal{L}{align}$ and global-uniformity $\mathcal{L}{uniform}$, which work together to generate unbiased representations. Finally, the interaction function utilizes these unbiased representations to predict scores for user-item pairs, $s(u,i)$, as part of the recommendation task $\mathcal{L}_{rec}$. It is important to note that while the diagram specifically focuses on item-side debiasing, similar operations are conducted on the user side as well.}
    \Description{An illustration of ~\shortname~framework.}
    \label{fig:AURL-model}
    \vspace{-0.5cm}
\end{figure}

We introduce a framework, \fullname, designed to mitigate biases in recommender systems through representation distribution.
The group-alignment module minimizes differences between popular and long-tail entities, thereby ensuring representation alignment. 
Inspired by contrastive learning~\cite{wang2020understanding}, our global-uniformity module aims to enhance the quality of representations.
The ~\shortname framework is illustrated in Fig. \ref{fig:AURL-model}. 
This section details the design and analysis of these modules, both theoretically and empirically, and presents the overall objective function.

\subsection{Group-Alignment module} 
In the previous section, we demonstrated that CF-based methods result in biased distributions in the latent space, termed "group-discrepancy". Consequently, despite similarities, items and users are dispersed based on popularity, leading to a model bias towards popular entities.
To address group-discrepancy, we propose group-alignment, aiming for popular and long-tail entities to share distribution characteristics. Due to data abundance and inherent user preference biases, popular entities' representations align more with user preferences, whereas long-tail entities' representations may be less precise. Our goal is to minimize the distributional distance between these entity groups in the representation space, thus standardizing representation distributions across groups. This involves transferring insights from well-established representations of popular entities to enhance those of long-tail entities. We formally define the optimality of group-alignment as follows:

\textbf{Definition}~(Perfect Group-Alignment). The representation distribution is perfect group-alignment if the distribution of the popular entity group $G_{pop}$ aligns perfectly with the distribution of the long-tail entity group $G_{tail}$, i.e., $p(f(G_{pop}))=p(f(G_{tail}))$.
Here $p(f(\cdot))$ represents the distribution of representations for different groups. Drawing inspiration from domain adaptation techniques, we aim to minimize the distributional distance between two groups, thereby achieving group-aligned representation distributions. The formal representation of this concept is as follows:
\begin{equation}
    \mathop{min}\limits_{\theta}\hat{D}(f(G_{pop}),f(G_{tail})),
\end{equation}
where $\hat{D}(\cdot,\cdot)$  is an estimate of the distribution discrepancy between the representation of popular user/item groups $f(G_{pop})$ and that of unpopular user/item groups $f(G_{tail})$.
Potential measures for this discrepancy include KL divergence~\cite{kiefer1952stochastic}, Maximum Mean Discrepancy (MMD)~\cite{Tolstikhin2016MinimaxEO}, among others.

Both KL divergence and Maximum Mean Discrepancy (MMD) are commonly used to measure the difference between two probability distributions.
However, KL divergence requires knowledge of the probability density functions involved and is sensitive to the specific forms and assumptions underlying these distributions~\cite{zhu2019aligning,kiefer1952stochastic}.
In contrast, MMD offers several advantages when aligning distributions. MMD effectively measures differences between two distributions without assuming specific forms~\cite{Tolstikhin2016MinimaxEO}. This is achieved by mapping distributions into a high-dimensional feature space and computing inner products between samples in this space to quantify differences. This approach captures overall distribution characteristics effectively, regardless of distribution shapes~\cite{rahman2020minimum}.
Furthermore, MMD eliminates the need for intermediate density estimation, resulting in more stable optimization and avoiding the challenges and instabilities associated with complex alternative optimization procedures~\cite{viehmann2021partial}. Additionally, by directly maximizing the difference between distributions, MMD reduces sensitivity to initialization. This guidance helps models learn better feature representations and avoid suboptimal solutions.
Overall, using MMD facilitates better alignment of different distributions and yields favorable outcomes in tasks like domain adaptation and unsupervised learning.

To achieve group alignment, we employ Maximum Mean Discrepancy (MMD)\cite{Tolstikhin2016MinimaxEO} as our regularizer to estimate the discrepancy between the two groups.
MMD functions as a kernel-based two-sample test that assesses the null hypothesis $p(f(G_{pop})) = p(f(G_{tail}))$, based on the observed samples, where the encoder $f(\cdot)$ maps users/items into the representation space~\cite{gretton2012kernel}.
The fundamental concept of MMD is that if the generating distributions are identical, then all statistical measures derived from these distributions should also be identical~\cite{zhu2019aligning}.
Formally, MMD quantifies the following difference measures:
\begin{equation}
    D_{\mathcal{H}}(G_{pop},G_{tail})=:\|\mathbb{E}[\phi(f(G_{pop}))]-\mathbb{E}[\phi(f(G_{tail}))]\|_{\mathcal{H}}^2,
\end{equation}
where $\mathcal{H}$ represents the reproducing kernel Hilbert space (RKHS) equipped with a characteristic kernel $k$~\cite{zhu2019aligning}.
Here, $\phi(\cdot)$ denotes the feature map of the original representation to RKHS, and the kernel $k$ is defined as $k(f(G_{pop}), f(G_{tail})) = <\phi(f(G_{pop})), \phi(f(G_{tail}))>$, where $<\cdot, \cdot>$ denotes the inner product of vectors.
The central theoretical result is that $p(f(G_{pop})) = p(f(G_{tail}))$ holds if and only if $D_{\mathcal{H}}(G_{pop},G_{tail}) = 0$~\cite{gretton2012kernel}.
In practice, an estimate of the MMD involves comparing the squared distance between the empirical kernel mean representations, denoted as:
\begin{equation}
    D_{\mathcal{H}}^{\backsim}(G_{pop},G_{tail})=:\|\frac{1}{| G_{pop}|}\sum_{x_i\in f(G_{pop})}\phi(x_i)-\frac{1}{| G_{tail}|}\sum_{x_j\in f(G_{tail})}\phi(x_j)\|_{\mathcal{H}}^2,\label{D unbias estimator}
\end{equation}
where $D_{\mathcal{H}}^{\backsim}(G_{pop},G_{tail})$ is an unbiased estimator of $D_{\mathcal{H}}(G_{pop},G_{tail})$~\cite{zhu2019aligning}. We use Eqn.~(\ref{D unbias estimator}) as the estimate of the discrepancy between $G_{pop}$ and $G_{tail}$.

Based on our previous discussion, both the user and item sides exhibit the issue of group-discrepancy. Therefore, we simultaneously apply MMD to both the user and item sides as follows:
\begin{equation}
\begin{aligned}
    \mathcal{L}_{align} &=\frac{1}{2} \times (\mathcal{L}_{align}^{user} + \mathcal{L}_{align}^{item}) \\
    &=\frac{1}{2}\times (D_{\mathcal{H}}^{\backsim}(f(G_{pop}^{user}),f(G_{tail}^{user}))+D_{\mathcal{H}}^{\backsim}(f(G_{pop}^{item}),f(G_{tail}^{item})))\\
    &=\frac{1}{2}\times (\|\frac{1}{| G_{pop}^{user}|}\sum_{x_i\in f(G_{pop}^{user})}\phi(x_i)-\frac{1}{| G_{tail}^{user}|}\sum_{x_j\in f(G_{tail}^{user})}\phi(x_j)\|_{\mathcal{H}}^2\\
    &+\|\frac{1}{| G_{pop}^{item}|}\sum_{x_i\in f(G_{pop}^{item})}\phi(x_i)-\frac{1}{| G_{tail}^{item}|}\sum_{x_j\in f(G_{tail}^{item})}\phi(x_j)\|_{\mathcal{H}}^2),
\end{aligned}
\label{equ:align}
\end{equation}
where $|\cdot|$ represents the size of the set.
It should be noted that the traditional training paradigm has effectively learned the representation of popular user/item groups. However, if the distance between the two group distributions is artificially reduced, there is a risk that the performance of the popular group may diminish.
Therefore, we fix the representations of the popular entities and unilaterally push the representation distribution of long-tail entities closer to that of popular entities.
As illustrated by the black arrow in Fig.~\ref{fig:AURL-model}, when we apply $\mathcal{L}_{align}$, the long-tail group (depicted with a blue dotted line) in the bias distribution moves closer to the popular group (depicted with an orange dotted line) in the representation space.
This operation allows the model to enhance the long-tail entity group without compromising the performance of the popular entity group.

\subsection{Global-Uniformity module}
In addressing the global-collapse issue, we draw inspiration from the uniformity property observed in contrastive learning~\cite{wang2020understanding} to develop a global-uniformity regularizer. This regularizer aims to enhance the even distribution of representation across various users and items, which we refer to as global-uniformity. Optimal global-uniformity means that feature vectors are spread as uniformly as possible across the unit hypersphere $\mathcal{S}^{m-1}$, thereby preserving a higher degree of informational content. We now proceed to define the criteria for optimal global-uniformity in the representation distributions within collaborative filtering as follows~\cite{Wang2022TowardsRA}:

\textbf{Definition}~(Perfect global-uniformity). The representation distribution is perfect global-uniformity if the distribution of $f (u) $  for $u \backsim p(\boldsymbol{U})$ and the distribution of $f(i)$ for $i\backsim p(\boldsymbol{I})$ are the uniform distribution $\sigma_{d-1} $ on $\mathcal{S}^{d-1}$.
Here, $\mathcal{S}^{d-1}=\{x\in \mathbb{R}_d:\|x\|=1\}$ represents the surface of the $d-dimenaional$ unit sphere, and $p(\boldsymbol{U})$ and $p(\boldsymbol{I})$ denote the distributions of users and items, respectively.
The global-uniformity property ensures that each representation preserves as much intrinsic information about the user or item as possible.
Studies have demonstrated that improved global-uniformity enhances the quality of representations both theoretically and empirically~\cite{wang2020understanding, Wang2022TowardsRA}.

To implement a global-uniformity optimizer, we aim for the global-uniformity property to be both asymptotically correct (i.e., the distribution optimized by this metric should converge to a uniform distribution) and empirically reasonable with a finite number of points, as described in~\cite{wang2020understanding}. To achieve this, we utilize the Gaussian potential kernel (also known as the Radial Basis Function (RBF) kernel), $G_t:\mathcal{S}^d\times\mathcal{S}^d\to\mathbb{R}_+$~\cite{borodachov2019discrete}:
\begin{equation}
    \setlength{\abovedisplayskip}{4pt}
    \setlength{\belowdisplayskip}{4pt}
    G_t(f(v),f(v^{'})) \triangleq e^{-t\|f(v)-f(v^{'})\|_2^2}=e^{2t\cdot f(v)^T f(v^{'})-2t},t>0,
\end{equation}
where $v$ and $v^{'}$ represent any user or item, and $t$ is a fixed parameter. We define the global-uniformity loss in the recommendation system as the logarithm of the average pairwise Gaussian potential on both the user side and item side:
\begin{equation}
    \setlength{\abovedisplayskip}{4pt}
    \setlength{\belowdisplayskip}{4pt}
    \begin{aligned}
        \mathcal{L}_{uniform}&=\frac{1}{2} \times (\mathcal{L}_{uniform}^{user}+\mathcal{L}_{uniform}^{item})\\ 
                        &=\frac{1}{2}\times (log\mathop{\mathbb{E}}\limits_{u,u^{'}\backsim p(\boldsymbol{U})}{G_t(u,u^{'})}+log\mathop{\mathbb{E}}\limits_{i,i^{'}\backsim p(\boldsymbol{I})}{G_t(i,i^{'})})\\
                        &=\frac{1}{2}\times (log\mathop{\mathbb{E}}\limits_{u,u^{'}\backsim p(\boldsymbol{U})}{e^{-t\|f(u)-f(u^{'})\|_2^2}}+log\mathop{\mathbb{E}}\limits_{i,i^{'}\backsim p(\boldsymbol{I})}{e^{-t\|f(i)-f(i^{'})\|_2^2}}).
    \end{aligned}
    \label{equ:uniform}
\end{equation}

It is worth noting that the perfectly uniform lower bound here is not a fixed value, but is dependent on the dimension of the representation space~\cite{wang2020understanding}.
The above formula demonstrates that $\mathcal{L}_{uniform}$ encourages users and items to be distributed as evenly as possible across the entire unit sphere. This promotes more effective utilization of semantic information within the space. Maintaining a uniform distribution of feature vectors maximizes the retention of the original feature information in the data.

\subsection{Analyses of Group-Alignment and Global-Uniformity}
\subsubsection{\textbf{Theoretical Analyses}}
We first explore the necessity of group-alignment in mitigating the item-side popularity bias. We define $\mathcal{L}_{debias}$ as the target for debiasing under ideal conditions, which aims to equalize the expected preferences of users towards both the popular and long-tail item groups.
Formally, $\mathcal{L}_{debias}$ for a user $u$ is defined as:
\begin{equation}
\begin{aligned}
    \mathop{min}\limits_{\theta} \mathcal{L}_{debias}&=\mathop{min}\limits_{\theta}\mathbb{E}_{i \in G_{pop}^{item}}[s(u,i)]-\mathbb{E}_{i \in G_{tail}^{item}}[s(u,I)]  \\
    &=\mathop{min}\limits_{\theta} \mathbb{E}_{i \in G_{pop}^{item}}[\sigma(\mathbf{z}_u^T f(i))]-\mathbb{E}_{i \in G_{tail}^{item}}[\sigma(\mathbf{z}_u^T f(i))]\\
    &=\mathop{min}\limits_{\theta} \sigma(\mathbf{z}_u^T(\mathbb{E}[f(G_{pop}^{item})]-\mathbb{E}[f(G_{tail}^{item})]))\\
    &\propto  \mathop{min}\limits_{\theta}  \mathbb{E}[f(G_{pop}^{item})-f(G_{tail}^{item})]\\
    &=\mathop{min}\limits_{\theta} \sum_{i \in G_{pop}^{item}} f(i)p(f(G_{pop}^{item})) - \sum_{i \in G_{tail}^{item}} f(i)p(f(G_{tail}^{item}))\\
    &=\mathop{min}\limits_{\theta} \sum_{i}f(i)[p(f(G_{pop}^{item})-p(f(G_{tail}^{item}))]\\
    \label{different}
\end{aligned}
\end{equation}
According to the definition of perfect group-alignment, Eqn.~(\ref{different}) tends towards zero only when the encoder achieves perfect group-alignment between the distributions of the two item groups. This suggests that reducing the distance between the distributions of the two item groups can help narrow the gap between the scores of the popular item group and the long-tail item group.

We then explore the necessity of group-alignment in mitigating the user-side consistency bias.
Let $\mathbb{E}[\mathcal{L}_{G_{pop}^{user}}]$ and $\mathbb{E}[\mathcal{L}_{G_{tail}^{user}}]$ represent the expected value of the loss for the popular user group and the long-tail user group, respectively. The Bayesian Personalized Ranking (BPR) loss can then be reformulated as:

\begin{equation}
\begin{aligned}
    \mathcal{L}_{BPR}&=\frac{1}{|\mathcal{R}|}\sum_{(u,i)\in \mathcal{R}, 
 i^- \in \boldsymbol{I}/\boldsymbol{I}^+_u} \mathcal{L}_{(u,i,i^-)}=\frac{1}{|\mathcal{R}|}\sum_{u \in |\boldsymbol{U}|}\sum_{i \in \boldsymbol{I}_u^+, i^- \in \boldsymbol{I} / \boldsymbol{I}_u^+}\mathcal{L}_{(u,i,i^-)}\\
    &=\frac{1}{|\mathcal{R}|}(\sum_{u \in G_{pop}^{user}}\sum_{i \in \boldsymbol{I}_u^+, i^- \in \boldsymbol{I} / \boldsymbol{I}_u^+}\mathcal{L}_{(u,i,i^-)}+\sum_{u \in G_{tail}^{user}}\sum_{i \in \boldsymbol{I}_u^+, i^- \in \boldsymbol{I} / \boldsymbol{I}_u^+}\mathcal{L}_{(u,i,i^-)})\\
    &=\frac{1}{|\mathcal{R}|}( |\mathcal{R}^+_{G_{pop}^{user}}| \times \mathbb{E}[\mathcal{L}_{G_{pop}^{user}}]+|\mathcal{R}^+_{G_{tail}^{user}}| \times \mathbb{E}[\mathcal{L}_{G_{tail}^{user}}])
\end{aligned}
\end{equation}
where $\mathcal{R}^+_{G_{pop}^{user}}$ and $\mathcal{R}^+_{G_{tail}^{user}}$ respectively represent the interaction sets of popular users and long-tail users in the training set.
Due to imbalanced data, the number of interactions for popular users ~($|\mathcal{R}^+_{G_{pop}^{user}}|$) significantly exceeds that of long-tail users ~($|\mathcal{R}^+_{G_{tail}^{user}}|$).
This imbalance causes the training process to optimize predominantly for popular users, achieving a lower average loss, while the impact on long-tail users remains minimal.
Consequently, the model tends to learn the characteristics and preferences of popular users more effectively, thereby optimizing the recommendation results for this group more successfully.
To enhance the accuracy for long-tail users, our objective is to balance the loss between the two user groups, expressed as $\mathbb{E}[\mathcal{L}_{G_{pop}^{user}}]-\mathbb{E}[\mathcal{L}_{G_{tail}^{user}}]$.
The proposed group-alignment property facilitates the transfer of knowledge from the well-trained representations of popular users to long-tail users, improving the representational distribution and learning of preferences for the latter.

Please note that achieving distribution consistency can be readily accomplished by encoding all items or users into the same representation; however, this approach risks inducing global-collapse~\cite{Wang2022TowardsRA}. To circumvent this issue, we have designed a global-uniformity regularizer. This regularizer ensures that representations are uniformly distributed across the unit hypersphere, thereby preserving maximal information about the entities.

\begin{figure}
    \centering
    \vspace{-0.5cm}
    \subfloat[NDCG@20]{\includegraphics[width =0.5\linewidth]{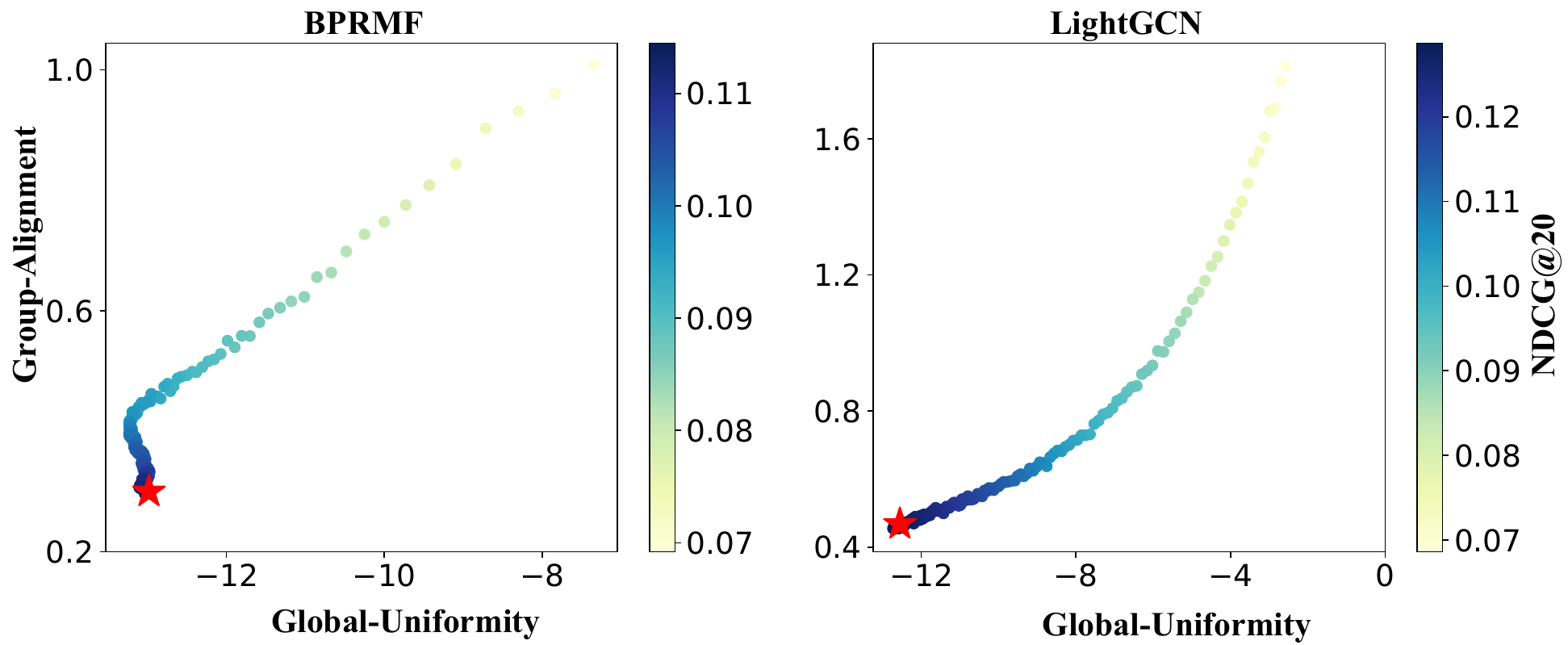}\label{fig:NDCG_AU}}
    \hfill
    \subfloat[PRU]{\includegraphics[width =0.5\linewidth]{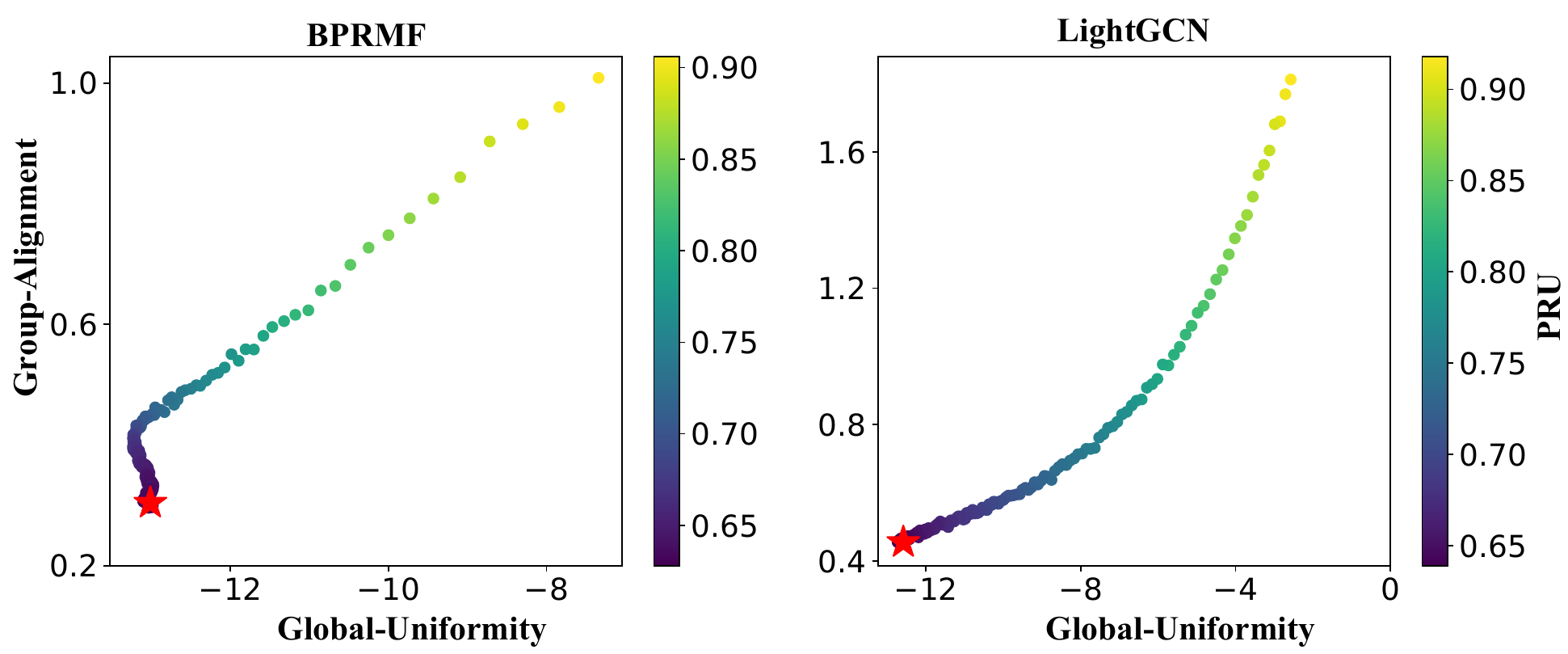}\label{fig:AU_PRU}}
    \vspace{-0.3cm}
    \caption{Metrics and performance of BPRMF and LightGCN experiments are visualized. Each point on the plot represents a trained encoder, with its \(x\) and \(y\) coordinates indicating group-alignment and global-uniformity properties, respectively. The color of each point denotes validation set accuracy, measured by $NDCG@20$, and debiasing metrics, represented by $PRU$. Stars on the plot indicate the points where convergence was achieved.}
    \label{fig:au-ndcg-pru}
    \Description{Metrics and performance of BPRMF and LightGCN experiments.}
    \vspace{-0.5cm}
\end{figure}

\subsubsection{\textbf{Empirical Observations}}
To further investigate the relationship between the two properties and recommendation results, we depicted the variations of the properties, along with the accuracy of recommendations ~($NDCG@20$) and the debias metrics ($PRU$), during the training process~(Fig.~\ref{fig:au-ndcg-pru}). Lower values of group-alignment and global-uniformity are desirable. A lower $PRU$ metric signifies less bias towards popularity in recommendations, while a higher $NDCG@20$ indicates better recommendation accuracy. The stars in Fig.~\ref{fig:au-ndcg-pru} represent the convergence points of the model.

We observe a highly consistent behavior between the properties and the recommendation metrics. The upper two subfigures illustrate the trends of $NDCG@20$ with group-alignment and global-uniformity, where darker colors represent better recommendation quality. It is noticeable that smaller values of global-uniformity and group-alignment are more favorable for $NDCG@20$ improvement.
The lower two subfigures in Fig.~\ref{fig:au-ndcg-pru} showcase the relationship between the debias indicator $PRU$ and representation distribution properties. Darker colors signify lower $PRU$ values, indicating better debiasing performance. As evident, achieving better group-alignment or global-uniformity can both help alleviate the biases of the recommendation results and improve the accuracy of the model, and it may be beneficial to optimize them simultaneously.

In summary, our theoretical and empirical analyses demonstrate a strong correlation between the ideal debiasing objectives and the two properties we proposed. In other words, better group-alignment and global-uniformity are advantageous for both improving the accuracy of the model's recommendations and mitigating bias.

\subsection{Objective function}
The loss function of our framework consists of three components: recommendation loss~($\mathcal{L}_{rec}$), group-alignment loss~($\mathcal{L}_{align}$), and global-uniformity loss~($\mathcal{L}_{uniform}$). 
For details, $\mathcal{L}_{rec}$ represents the loss function for the recommendation task in any CF-based model. This could be, for example, the BPR loss~\cite{Rendle2009BPRBP}, InfoNCE loss~\cite{Yu2021AreGA}, Softmax loss~\cite{citationsurveylekey}, or any other suitable loss function used in collaborative filtering models.
To prevent overfitting, we also apply $L_2$ regularization to constrain all the parameters of the model.
The overall loss function is formulated as follows:

\begin{equation}
\mathcal{L}_{\shortname}=\mathcal{L}_{rec}+\lambda_{1}\mathcal{L}_{align}+\lambda_{2}\mathcal{L}_{uniform}+\lambda\|\theta\|_F^2,
\label{loss function}
\end{equation}
where $\lambda_{1}$ and $\lambda_{2}$ are hyperparameters that control the strengths of auxiliary tasks, $\theta=\{\mathbf{Z},\mathbf{H}\}$ denotes all trainable model parameters, and $\lambda$ governs the $L_2$ regularization strength to prevent overfitting. Additionally, it is important to note that the auxiliary tasks do not introduce additional parameters. 
In our framework, by jointly training $\mathcal{L}_{align}$ and $\mathcal{L}_{uniform}$, we achieve unbiased representations that help debias recommendation results.
~\shortname~ is designed to alleviate biases in results from both the item side and the user side. 
Furthermore, ~\shortname~ is compatible with arbitrary CF-based models by integrating two auxiliary loss functions.

\subsection{Discussion}
In this work, we primarily investigate the issue of popularity bias in recommendation system results and propose the ~\fullname ~. We address both item-side popularity bias and user-side consistency bias by regularizing the distribution of representations, detailing the model of ~\shortname. Specifically, we elaborate on the mechanism behind the group alignment regularizer, which aims to mitigate group discrepancies by aligning the distribution of long-tail entities with that of popular entities in the representation space. Additionally, we discuss the global uniformity regularizer, which combats global collapse phenomena by encouraging representations of entities on the hypersphere to be more uniformly distributed.

While our method has shown some effectiveness in addressing bias in recommendation systems, it has limitations. Our approach, which involves aligning and uniformizing the entire item/user space, might not effectively capture subtle differences, highlighting the need for more fine-grained grouping methods. Additionally, despite achieving group alignment, the method's underlying mechanism lacks intuitiveness and interpretability, which could hinder understanding and communication with users and stakeholders. These aspects suggest significant areas for further research and improvement in the interpretability of the method.

\section{Experiments}

\subsection{Experimental settings}
\subsubsection{\textbf{Dataset}.} 
We conduct experiments on three public real-world datasets: Amazon-Book for e-commerce recommendation~\cite{Yu2021AreGA}, Movielens-20M for movie recommendation~\cite{harper2015movielens}, and Douban-Book for book recommendation~\cite{Yu2021AreGA}. Following previous experimental setups~\cite{Yu2021AreGA, He2020LightGCNSA}, we retain only users and items with at least 5 interactions. For each user, 70\% of interactions are randomly selected as the training set, another 10\% as the validation set for hyperparameter tuning, and the remaining 20\% as the test set.

To define user and item groups, we apply the "Pareto Principle", selecting the top 20\% most frequently interacted users/items in the training set as the popular group, and the others as the long-tail group~\cite{perc2014matthew}. Detailed data statistics are summarized in Table~\ref{tab:stats}.

\begin{table}
    \centering
    \caption{Detailed datasets statistics.}
    \vspace{-0.2cm}
    \resizebox{0.6\columnwidth}{!}{
    \begin{tabular}{ccccl}
    \toprule
   \textbf{ Datasets}&\textbf{\#Users}&\textbf{\#Itmes}&\textbf{\#Interactions}&\textbf{Density}\\
    \midrule
    \textbf{Amazon-Book}&52,643&91,599&2,984,108&0.0619\%\\
    \textbf{Movielens-20M}&99,626&14,387&28,011,110&0.1954\%\\
    \textbf{Douban-Book}&12,859&22,294&792,062&0.2087\%\\
    \bottomrule
    \end{tabular}}
    \label{tab:stats}
    \vspace{-0.5cm}
\end{table}

\subsubsection{\textbf{Evaluation metrics.}}
In this work, we place greater emphasis on the biases in model results rather than on accuracy. Therefore, we adopt two accuracy metrics to ensure that ~\shortname~ does not significantly compromise the model's accuracy, and we introduce two metrics to demonstrate how effectively ~\shortname~ mitigates biases.
To measure the accuracy of the recommendation results, we utilize Hit Ratio ~($HR@K$)~ and Normalized Discounted Cumulative Gain ~($NDCG@K$)~\cite{He2020LightGCNSA}. To evaluate the debiasing capability of the model, we assess it from the perspectives of both users and items.

For the item side, we select Popularity-Rank Correlation ($PRU$)~\cite{Zhu2021PopularityOpportunityBI, Zhao2022InvestigatingAP}, which measures the correlation between the user's true preference and the item's popularity. Specifically, $PRU$ is defined as:
\begin{equation}
    PRU=-\frac{1}{|\textbf{\textit{U}}|}\sum_{u\in \textbf{\textit{U}}}SRC(pop(\boldsymbol{O^+_u}),rank(\boldsymbol{O^+_u})),
\end{equation}
where $SRC(\cdot)$ is the Spearman Rank Correlation coefficient, and $\boldsymbol{O^+_u}$ represents the set of items interacted with by user $u$ in the test dataset, reflecting user preferences. In particular, a smaller value of $PRU$ indicates that the model more accurately captures the user's true preference, independent of item popularity.

For the user side, we employ Demographic Parity ~($DP@K$) to evaluate the consistency of recommendation results among different user groups~\cite{fairdata23, fairgo, Chen2020BiasAD}. This metric utilizes Jensen–Shannon Divergence, $JSD(\cdot,\cdot)$, to measure the distributional distance in accuracy among different groups, specifically:
\begin{equation}
    DP@K=JSD(ACC(G_{pop}^{user}),ACC(G_{tail}^{user})),
\end{equation}
where $ACC(G_{pop}^{user})$ and $ACC(G_{tail}^{user})$ represent the distribution of recommendation accuracy for the popular user group and long-tail user group, respectively. A smaller value of $DP@K$ indicates less bias on the user side. 

\subsubsection{\textbf{Baselines.}}
We implement ~\shortname~ using three classic CF-based models as the backbone to validate its effectiveness. The three backbones are:

\begin{enumerate}
    \item \textbf{BPRMF}~\cite{Rendle2009BPRBP}: Maps user/item IDs into the representation space using matrix factorization techniques, capturing intricate relationships between users and items.
    \item \textbf{LightGCN}~\cite{He2020LightGCNSA}: An advanced CF-based model based on GCNs. It captures higher-order interactions between users and items.
    \item \textbf{SimGCL}~\cite{Yu2021AreGA}: A cutting-edge Graph Contrastive Learning (GCL) method. It adds uniform noise to node representations when generating different views.
\end{enumerate}
We compare our methods with various state-of-the-art models in three broad categories, as follows:
\begin{itemize}[leftmargin=0cm, itemindent=0.3cm]
    \item [$\bullet$] \textbf{Re-weighting-based methods:}
    \begin{itemize} 
        \item[] (1) \textbf{RS~\cite{rendle2014improving}}. The key to RS is to balance the number of interactions for high item/user and low engagement training by resampling. 
        \item[] (2) \textbf{PC~\cite{Zhu2021PopularityOpportunityBI}}. This is a post-processing approach that directly modifies the prediction score by compensating based on the item/user popularity. 
        \item[] (3) \textbf{Zerosum~\cite{rhee2022countering}}. It reduces model bias in recommendation systems by directly equalizing recommendation scores across items preferred by a user.
        \item[] (4) \textbf{r-AdjNorm~\cite{Zhao2022InvestigatingAP}}. It obtains asymmetric aggregation by adjusting the gamma parameter during aggregation in the graph so that the model is more inclined to long-tail nodes. 
        \item[] (5) \textbf{CPFair~\cite{naghiaei2022cpfair}}. It integrates fairness constraints from both consumer and producer perspectives in a joint objective framework for recommender systems based on the re-ranking approach.
    \end{itemize}
    
    \item[$\bullet$] \textbf{Decorrelation-based methods:}
    \begin{itemize}
        \item[] (6) \textbf{DICE~\cite{zheng2021disentangling}}. Based on the idea of decoupling, it designs a causal data framework that decomposes user interest and item popularity into two representations. 
        \item[] (7) \textbf{MACR~\cite{Wei2020ModelAgnosticCR}}. MACR is a method based on counterfactual reasoning for eliminating the item side bias in the result.
    \end{itemize}
    \item[$\bullet$] \textbf{Adversarial-based methods:}
    \begin{itemize}
        \item[] (8) \textbf{FairMI~\cite{zhao2023fair}}. It leverages adversarial principles to minimize mutual information between embeddings and sensitive attributes, while simultaneously maximizing it with non-sensitive information.
        \item[] (9) \textbf{InvCF~\cite{zhang2023invariant}}. It exploits the notion that item/user representations remain unchanged despite variations in popularity semantics. By filtering out unstable or outdated popularity characteristics, InvCF learns unbiased representations.
    \end{itemize}
\end{itemize}

In summary, the baselines cover a wide array of methods aimed at debiasing recommendation results. These techniques include re-sampling, re-weighting, regularization, re-ranking, causal approaches, among others.
For a fair comparison, we exclude methods that necessitate additional training data, such as Autodebias~\cite{AutoDebias} and DecRs~\cite{DecRS}.

\subsubsection{\textbf{Hyper-Parameter Settings.}}
In all our experiments, the number of negative samples is set to 1, consistent with other studies ~\cite{Yu2021AreGA,He2020LightGCNSA,Zhu2021PopularityOpportunityBI}. We employ the Xavier initializer~\cite{glorot2010understanding} to set up the parameters and utilize Adam~\cite{DBLP:journals/corr/KingmaB14} with a learning rate of 0.001 for optimizing all models. The representation size is fixed at 64, the batch size at 2048, and the $L_2$ regularization coefficient at 0.0001. For all baseline hyper-parameters, we adopt the values recommended in the respective papers and meticulously adjust them for the new dataset to optimize recommendation outcomes. We fine-tune $\lambda_1$ and $\lambda_2$ over the range { $10^{-6}$, $10^{-5}$, $10^{-4}$, $10^{-3}$, $10^{-2}$, $10^{-1}$, 1} with an appropriate step size.
We run all models five times and report the mean results.

\subsection{Overall performance}
\begin{figure}[t]
    \centering
    \vspace{-0.2cm}
    \subfloat[Item side] {\includegraphics[width=0.4\linewidth]{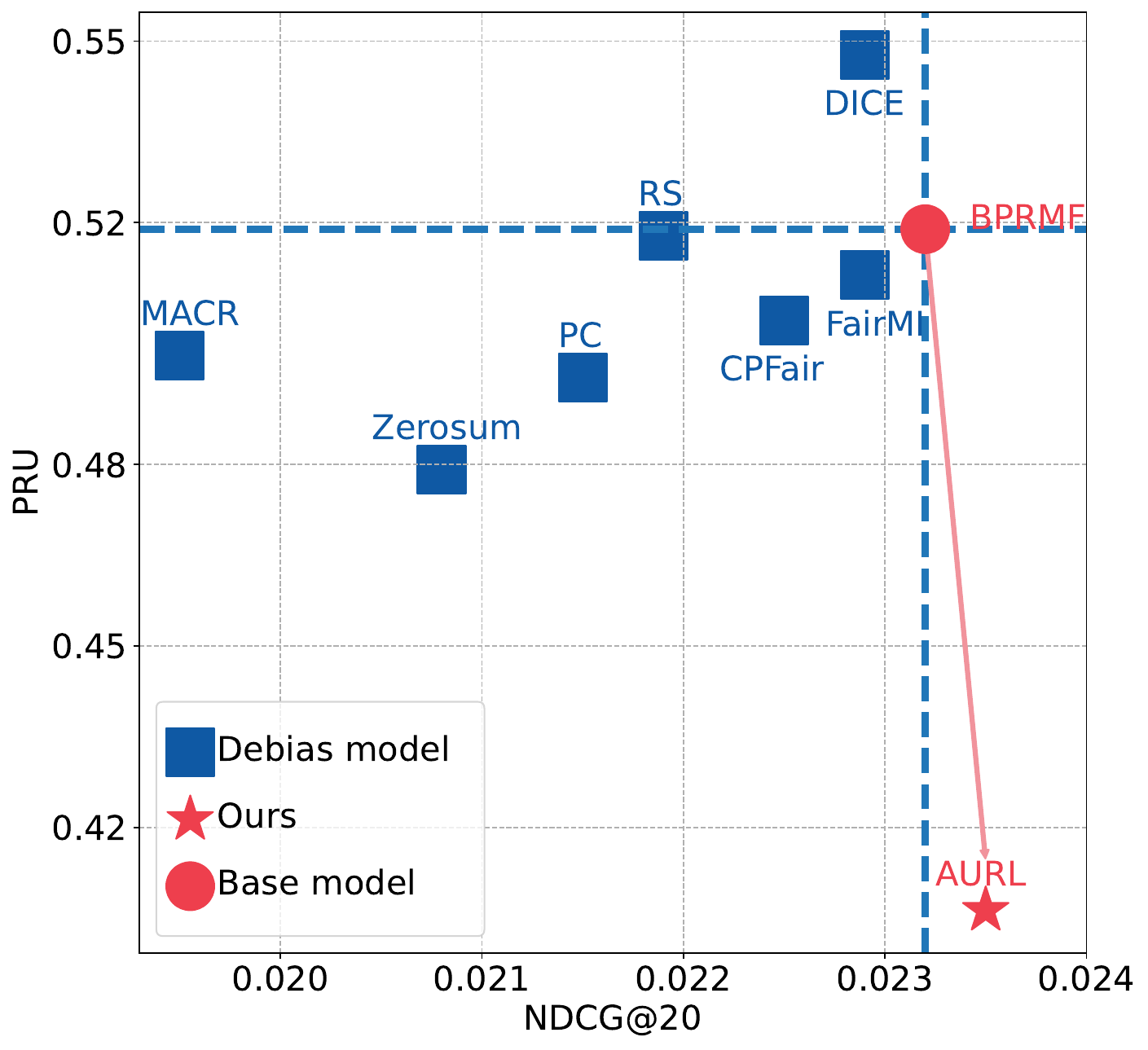}\label{NDCG@20-PRU}}
    \subfloat[User side] {\includegraphics[width=0.4\linewidth]{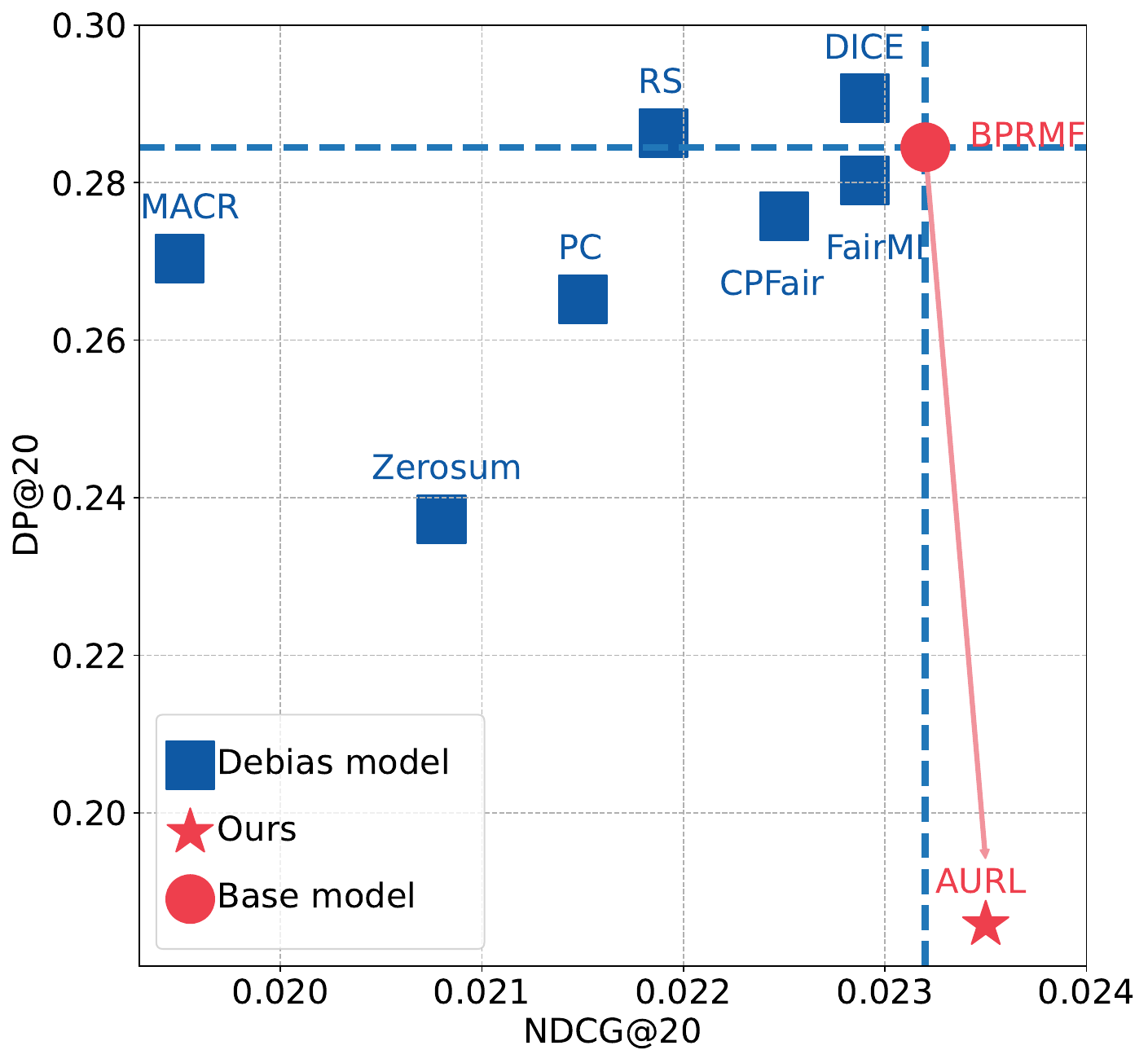}\label{NDCG@20-DP@20}}
    \vspace{-0.2cm}
    \caption{Trade-off between recommendation accuracy and debiasing efforts in the Amazon-Book dataset.}
    \label{fig:model}
    \Description{Trade-off between recommendation accuracy and debiasing on Amazon-Book dataset.}
    \vspace{-0.5cm}
\end{figure}

As demonstrated in Table~\ref{bprmf overall}, Table~\ref{tab: LightGCN overrall}, and Table~\ref{tab: SimGCL overrall}, we compare ~\shortname ~ with various debiasing methods across different backbones to assess the overall performance.
To better elucidate the trade-off between recommendation accuracy and debiasing efforts, we visualize the relationships among recommendation accuracy ($NDCG@20$), item-side debiasing ($PRU$), and user-side debiasing ($DP@20$) across various models, as illustrated in Fig.~\ref{fig:model}.
Based on the experimental results, we observe the following key points.

\begin{table}[t]
    \caption{Performance of the ~\shortname ~ variant and the baseline using the BPRMF backbone with $K=20$, where $\uparrow$ indicates that higher values are preferable and $\downarrow$ indicates that lower values are preferable.}
    \vspace{-0.2cm}
    \label{bprmf overall}
    \resizebox{1\columnwidth}{!}{
    \begin{tabular}{|c|c|c|c|c|c|c|c|c|c|c|c|c|}
    \hline
    \multirow{2}*{\textbf{Model}} & \multicolumn{4}{c|}{\textbf{Amazon-Book}}     & \multicolumn{4}{c|}{\textbf{Movielens-20M}} & \multicolumn{4}{c|}{\textbf{Douban-Book}}     \\
    \cline{2-13}
     & \textbf{HR$\uparrow$} &\textbf{ NDCG$\uparrow$ }& \textbf{PRU$\downarrow$ }& \textbf{DP$\downarrow$ }& \textbf{HR$\uparrow$} & \textbf{NDCG$\uparrow$} & \textbf{PRU$\downarrow$ }&\textbf{ DP$\downarrow$ } &\textbf{ HR$\uparrow$} & \textbf{NDCG$\uparrow$} & \textbf{PRU$\downarrow$} & \textbf{DP$\downarrow$}\\
    \hline
   \textbf{ BPRMF }&\underline{0.0318} &\underline{0.0232}&0.5189&0.2845& 
   \underline{0.3291}&0.2284  &0.6400 &0.2352 &
    \textbf{0.1290}&\underline{ 0.1027}&0.6692&0.2923\\
    \hline
    \textbf{+RS}&0.0297&0.0219&0.5178&0.2863&
    0.2971&\underline{0.2297}&0.6242&0.1946&    
    0.1151&0.0937&0.6065&0.2136\\
    
    \textbf{+PC}&0.0297&0.0215&0.4944&0.2652&
    0.3269&0.2274&\underline{0.6208}&0.2451&          
    0.1174&0.0977&\underline{0.5840}&0.1704\\
    \textbf{+Zerosum}&0.0287&0.0208&\underline{0.4792}&\underline{0.2373}&
    0.3057&0.2141&0.6302&\underline{0.1772}&      
    0.1115&0.0905&0.6436&0.2029\\
    \textbf{+CPFair}&0.0309 & 0.0225 & 0.5038 & 0.2757 & 0.3194 & 0.2215 & 0.6208 & 0.2280 & 0.1250 & 0.0995 & 0.6490 & 0.2837\\
    \hline
    \textbf{+DICE}&0.0310&0.0229&0.5477&0.2907&
    0.3204&0.2282&0.6358&0.2392&
    0.1213&0.1001&0.6106&\underline{0.1641}\\
    \textbf{+MACR}&0.0297&0.0195&0.4980&0.2704&
    0.3180&0.2219&0.6299&0.2019&
    0.1030&0.0849&0.6607&0.1845\\
    
    \hline
    \textbf{+InvCF}&0.0316 & 0.0230 & 0.5241 & 0.2873 & 0.3264 & 0.2261  & 0.6464 & 0.2377 &  0.1284 & 0.1016  & 0.6759 & 0.2955\\
    \textbf{+FairMI}&0.0314 & 0.0229 & 0.5113 & 0.2803 & 0.3243 & 0.2251 & 0.6305 & 0.2318 & 0.1274 & 0.1013 & 0.6592 & 0.2880\\
    \hline
    \textbf{+~\shortname}& \textbf{0.0322} & \textbf{0.0235}& \textbf{0.4063}&\textbf{0.1858}&
    \textbf{0.3372}&\textbf{0.2370}&\textbf{0.5168}&\textbf{0.1341}&
    0.1232&\textbf{0.1045} &\textbf{0.4814}&\textbf{0.1850}\\

   \textbf{ Improvement(\%)}& \textbf{1.25\%}&\textbf{1.29\%}&\textbf{15.21\%}&\textbf{21.70\%}&
    \textbf{2.46\%}&\textbf{3.17\%}&\textbf{16.75\%}&\textbf{24.32\%}&
    \textbf{-4.59\%}&\textbf{2.75\%}&\textbf{17.57\%}&\textbf{12.74\%}
\\
    \hline
    \end{tabular}}
    \vspace{-0.5cm}
\end{table}

\begin{table}[t]
    \caption{Performance of the ~\shortname ~ variant and the baseline using the LightGCN backbone with $K=20$, where $\uparrow$ indicates that higher values are preferable and $\downarrow$ indicates that lower values are preferable.}
    \vspace{-0.1cm}
    \label{tab: LightGCN overrall}
    \resizebox{1\columnwidth}{!}{
    \begin{tabular}{|c|c|c|c|c|c|c|c|c|c|c|c|c|}
    \hline
    \multirow{2}*{\textbf{Model}} & \multicolumn{4}{c|}{\textbf{Amazon-Book}}     & \multicolumn{4}{c|}{\textbf{Movielens-20M}} & \multicolumn{4}{c|}{\textbf{Douban-Book}}     \\
    \cline{2-13}
      & \textbf{HR$\uparrow$} &\textbf{ NDCG$\uparrow$ }& \textbf{PRU$\downarrow$ }& \textbf{DP$\downarrow$ }& \textbf{HR$\uparrow$} & \textbf{NDCG$\uparrow$} & \textbf{PRU$\downarrow$ }&\textbf{ DP$\downarrow$ } &\textbf{ HR$\uparrow$} & \textbf{NDCG$\uparrow$} & \textbf{PRU$\downarrow$} & \textbf{DP$\downarrow$}\\
    \hline
    \textbf{LightGCN }&0.0380 & 0.0282&0.4955 &0.2796 &
    0.3145 & 0.2172& 0.3453&0.2146 &
    0.1550&0.1282 &0.6319& 0.2804 \\
    \hline
    \textbf{+RS}&0.0386&0.0280&0.4934&0.2803&
    0.3085&0.2130&0.3408&0.2149&  
    0.1553&   0.1285     &   0.5905  &0.2606\\
    \textbf{+PC}&0.0378&0.0282&0.4921&0.2703&
     0.3085&0.2130&0.3348&0.2139&          
    0.1505& 0.1230  &   0.5899  &0.2646\\
    \textbf{+Zerosum}&0.0376&0.0274&0.4911&\underline{0.2736}&
    0.3031&0.2087&\underline{0.3342}&\underline{0.2044}&     
    0.1534&   0.1256  &   \underline{0.5719}    &\underline{0.2536}\\
    \textbf{+r-AdjNorm}&0.0387&0.0288&\underline{0.4901}&0.2671&
    0.3214&0.2199&0.4361&0.2255& 
    0.1664& \underline{0.1401}&0.6196&0.2687\\
    \textbf{+CPFair}&0.0370&0.0274&0.4950&0.2743&0.3054 & 0.2108 & 0.3360 & 0.2090 & 0.1525 & 0.1244 & 0.6137 & 0.2721\\
    \hline
    \textbf{+DICE}&0.0384&0.0284&0.4925&0.2772&
    0.3179&0.2164&0.3402&0.2083&
    0.1562&0.1285&0.6168&0.2747\\
    \textbf{+MACR}&0.0356&0.0264&0.4951&0.3283&
    0.2905&0.1974&0.3442&0.2144&       
    0.1392&0.1028&0.6230&0.2792\\

    \hline
    \textbf{+InvCF}&\underline{0.0409}&\underline{0.0309}&0.5028&0.2691&\underline{0.3241}&\underline{0.2207}&0.4846&0.2247&\underline{0.1690}&0.1388&0.6260&0.2656\\
    \textbf{+FairMI}&0.0391 &0.0291 & 0.4906 & 0.2761 &0.3244 & 0.2237 &0.3408 & 0.2123 & 0.1597 & 0.1322 &0.6259 & 0.2772\\
    \hline
    \textbf{+\shortname}&\textbf{0.0458}&\textbf{0.0349}&\textbf{0.3543}&\textbf{0.2399}&
    \textbf{0.3495}&\textbf{0.2450}&\textbf{0.2992}&\textbf{0.1884}&
    \textbf{0.1831}&\textbf{0.1585}&\textbf{0.3599}&\textbf{0.2247}\\
    \textbf{Improvement(\%)}&\textbf{11.98\%}&\textbf{12.94\%}&\textbf{27.71\%}&\textbf{8.99\%}&
    \textbf{7.84\%}&\textbf{11.01\%}&\textbf{26.69\%}&\textbf{7.82\%}&
    \textbf{8.34\%}&\textbf{13.13\%}&\textbf{37.06\%}&\textbf{11.40\%}\\
    \hline
    \end{tabular}}
    \vspace{-0.2cm}
\end{table}

\begin{table}[t]
    \caption{Performance of the ~\shortname ~ variant and the baseline using the SimGCL backbone with $K=20$, where $\uparrow$ indicates that higher values are preferable and $\downarrow$ indicates that lower values are preferable.}
    \vspace{-0.2cm}
    \label{tab: SimGCL overrall}
    \resizebox{1\columnwidth}{!}{
    \begin{tabular}{|c|c|c|c|c|c|c|c|c|c|c|c|c|}
    \hline
     \multirow{2}*{\textbf{Model}} & \multicolumn{4}{c|}{\textbf{Amazon-Book}}     & \multicolumn{4}{c|}{\textbf{Movielens-20M}} & \multicolumn{4}{c|}{\textbf{Douban-Book}}     \\
    \cline{2-13}
      & \textbf{HR$\uparrow$} &\textbf{ NDCG$\uparrow$ }& \textbf{PRU$\downarrow$ }& \textbf{DP$\downarrow$ }& \textbf{HR$\uparrow$} & \textbf{NDCG$\uparrow$} & \textbf{PRU$\downarrow$ }&\textbf{ DP$\downarrow$ } &\textbf{ HR$\uparrow$} & \textbf{NDCG$\uparrow$} & \textbf{PRU$\downarrow$} & \textbf{DP$\downarrow$}\\
    \hline
    \textbf{SimGCL} &0.0537 &0.0414 &0.2154 &0.2500 &
    \underline{0.3603}& \underline{0.2624}&0.4575 &0.1751 &
    \textbf{0.1927}& \textbf{0.1677} &0.3653&0.2401  \\
    \hline
    \textbf{+RS}&0.0517&0.0402&0.2142&0.2592&
    0.3583&0.2619&0.4551&0.1740&    
   \underline{0.1925} &   \underline{0.1673}      &  0.3604    &\underline{0.2330}\\
   \textbf{+PC}&0.0537&0.0414&0.2133&0.2455&
    0.3596&0.2532&0.4340&0.1906&          
    0.1862&  0.1634 &  0.3656   &0.2416\\
    \textbf{+Zerosum}&0.0506&0.0403&0.2136&\underline{0.2438}&
    0.3485&0.2508&0.4027&0.1692&      
    0.1865&    0.1570 &  0.3450    &0.2378\\
    \textbf{ +r-AdjNorm}& \textbf{0.0551}&\textbf{0.0426}&\underline{0.2113}&0.2736&
  0.3518 &0.2573&\underline{0.3965}&\underline{0.1629}&       
    0.1875&     0.1631  &\underline{0.3329}&0.2675\\
    \textbf{+CPFair}& 0.0521&0.0400&0.2186& 0.2445&0.3492& 0.2551& 0.4481 & 0.1710 & 0.1901 & 0.1640& 0.3562 & 0.2377\\
    \hline
    \textbf{+DICE}&0.0526&0.0404&0.2193&0.2673&
    0.3603&0.2606&0.4562&0.1764&
   0.1921 &0.1635&0.3546&0.2361\\
    \textbf{+MACR}&0.0508&0.0397&0.2310&0.2858&
    0.3359&0.2474&0.5286&0.1770&       
    0.1684&  0.1490     &0.3951&0.2792\\ 
    \hline
    \textbf{+InvCF}&0.0540&0.0411&0.2486&0.2847&0.3402&0.2495&0.4620&0.1768&0.1846&0.1539&0.3847&0.2730\\
    \textbf{+FairMI}&0.0537&0.0409&0.2149&0.2482&0.3592&0.2597&0.4183&0.1667&0.1924&0.1669&0.3492&0.2342\\
    \hline
    \textbf{+\shortname}&\underline{0.0541}&\underline{0.0418}&\textbf{0.1875}&\textbf{0.2281} &
    \textbf{0.3616}&\textbf{0.2632}&\textbf{0.3638}&\textbf{0.1559}&
    0.1884&0.1644&\textbf{0.2967}&\textbf{0.2192}\\
    
    \textbf{Improvement(\%)}&\textbf{-1.98\%}&\textbf{-1.72\%}&\textbf{11.27\%}&\textbf{6.31\%}&
   \textbf{0.33\%}&\textbf{0.30\%}&\textbf{10.76\%}&\textbf{4.29\%}&
    \textbf{-2.23\%}&\textbf{-1.96\%}&\textbf{10.87\%}&\textbf{5.92\%}\\
    \hline
    \end{tabular}}
    \vspace{-0.5cm}
\end{table}

\begin{itemize}
    \item [$\bullet$] Traditional debiasing methods, such as re-weighting, decorrelation, and adversarial approaches, have shown promise in reducing bias in recommendation results. These methods typically surpass backbone models in metrics like $PRU$ and $DP$, indicating their potential to partially mitigate biases.
    However, these methods tend to overly focus on reducing the over-representation of popular users/items. As a result, the performance on these popular entities is deliberately decreased, leading to a notable decline in overall model performance.
    For example, using BPRMF as the backbone, among all traditional debiasing methods, Zerosum exhibits a relative decrease in metrics like $NDCG@20$ and $HR@20$. Nonetheless, it achieves the best performance in terms of $PRU$ and $DP@20$ across various datasets.
    Beyond traditional techniques, we also explore state-of-the-art methods like r-adjnorm, which adjusts standardization terms in graph aggregation.
    We have discovered that for graph-based models, fine-tuning aggregation parameters yields the most effective debiasing results. This improvement is largely due to the ability of these parameters in GCNs to balance the influence of long-tail nodes, thus enhancing both the debiasing effect and the model's accuracy.

    \item [$\bullet$] \shortname ~ significantly outperforms existing methods in reducing biases across all datasets and backbone models. For example, compared to the best existing debiasing baselines, our method shows remarkable performance improvements on the BPRMF backbone across three real datasets, with gains of 15.21\%, 21.70\%, 16.75\%, 24.32\%, 17.57\%, and 12.74\% in the $PRU$ and $DP@20$ metrics. Similarly, under the LightGCN backbone, improvements include 27.71\%, 8.99\%, 26.69\%, 7.82\%, 37.06\%, and 11.40\%. Additionally, with the SimGCL backbone, our approach enhances performance by 11.27\%, 6.31\%, 10.76\%, 4.29\%, 10.87\%, and 5.92\%. These enhancements are the result of optimizing two critical properties in the representation space: group-alignment and global-uniformity.
    
    \item [$\bullet$] \shortname ~ significantly reduces biases in recommendation results while also preserving model accuracy to a notable degree, thus enhancing the model's practical utility. For instance, with BPRMF as the backbone, ~\shortname ~ achieves performance gains in $HR@20$ and $NDCG@20$ across three datasets, with improvements of 1.25\%, 1.29\%, 2.46\%, 3.17\%, -4.59\%, and 2.75\% respectively. With LightGCN as the backbone, these gains are even more pronounced, with increases of 11.98\%, 12.94\%, 7.84\%, 11.01\%, 8.34\%, and 13.13\%. These improvements demonstrate that ~\shortname ~ effectively mitigates bias without sacrificing accuracy, making it highly suitable for practical applications, as depicted in Fig.~\ref{fig:model}, where it is clear that other methods compromise accuracy to reduce bias. Additionally, ~\shortname ~ shows significantly better performance enhancements compared to other graph-based backbone models. This can be attributed to the inherent properties of graph convolution operations, where nodes assimilate information from their neighbors. Through successive convolution layers, node representations tend to converge, reducing their discriminative power between different entities and resulting in a less uniform distribution of representations in graph-based collaborative filtering methods.
    
    \item [$\bullet$] 
    GCL-based methods like SimGCL optimize the uniformity of representations, enhancing their informativeness to a degree. Consequently, ~\shortname ~ does not significantly improve accuracy beyond what is achieved by SimGCL alone. However, the combination of SimGCL and ~\shortname ~ consistently and substantially reduces biases. Specifically, when compared with the strongest baseline methods, ~\shortname ~ achieves notable improvements in $PRU$ and $DP@20$, with increases of 11.27\% and 6.31\% on the Amazon-Book dataset, 10.76\% and 4.29\% on Movielens-20M, and 10.87\% and 5.92\% on Douban-Book, respectively. These results underscore the significant impact of our proposed group-alignment approach in mitigating biases.
\end{itemize}

\begin{table}[t]
    \centering
    \caption{Effect of different model components on BPRMF+\shortname.}
    \vspace{-0.2cm}
    \label{tab:ablation study}
    \resizebox{0.45\columnwidth}{!}{
    \begin{tabular}{ccccc}
        \hline
      \multirow{2}*{\textbf{Model}} & \multicolumn{4}{c}{\textbf{Douban-book}}     \\
    \cline{2-5}
      & \textbf{HR$\uparrow$} & \textbf{NDCG$\uparrow$ }& \textbf{PRU$\downarrow$} & \textbf{DP$\downarrow$}\\
        \hline
        \textbf{BPRMF}&

    0.1290& 0.1027&0.6692&0.2923\\
        \hline    
        \textbf{\shortname w/o AL}&\textbf{0.1325}&0.1031&0.6231&0.2403\\
        \textbf{\shortname w/o UN}&0.1213&0.1001&0.6116&0.1642\\
        \hline
        \textbf{\shortname w/o U}&0.1261&0.0993&0.4872&0.2623\\
        \textbf{\shortname w/o I}&0.1236&\textbf{0.1046}&0.5571&0.1847\\     
        \hline
        \textbf{BPRMF+\shortname}&0.1295&0.1027&\textbf{0.4070}&\textbf{0.1558}\\
        \hline
        \end{tabular}}
        \vspace{-0.5cm}
\end{table}

\subsection{Ablation study}
To demonstrate the effectiveness of the various components of ~\shortname, we conducted ablation studies on the Douban-Book dataset using BPRMF+\shortname:
(1)  \textbf{~\shortname ~ w/o AL}: This variant removes the group-alignment module to examine its impact on the consistency of representation distribution among different groups.
(2) \textbf{~\shortname ~ w/o UN}: This configuration excludes the global-uniformity module to assess the influence of the global-uniformity property on the model.
(3) \textbf{~\shortname ~ w/o U} and \textbf{~\shortname ~ w/o I}: These versions respectively remove the regularization terms on the user and item sides, testing the model’s ability to concurrently mitigate biases for both user and item sides.
By comparing these variants with the complete ~\shortname ~ model, we gain insights into the specific contributions of each component and assess their impact on the overall model performance. The results of this study are presented in Table~\ref{tab:ablation study}.

Analyzing the results of ~\shortname ~ w/o AL and ~\shortname ~ w/o UN, we observe that these modules enhance performance with respect to $PRU$ by 6.89\% and 8.61\%, respectively. This indicates that both components effectively alleviate bias in the model.
Additionally, while the group-alignment module significantly mitigates bias, it also results in a slight decrease in recommendation accuracy. In contrast, the global-uniformity module is vital for maintaining the accuracy of the recommendation model. The synergistic interaction between these two components not only ensures the accuracy of recommendations but also effectively addresses the issue of bias.

After comparing the experimental results of ~\shortname ~ w/o U and ~\shortname ~w/o I, it is evident that addressing the distribution of either users or items separately can effectively alleviate biases in recommendation results. This demonstrates that biases are prevalent on both the user and item sides, underscoring the necessity and effectiveness of ~\shortname ~ in treating these biases within a unified framework.
Specifically, when item representations are treated separately (~\shortname ~ w/o U), there is a more significant improvement in $PRU$, with an increase of 27.19\%. Conversely, addressing the user side alone (~\shortname ~w/o I) leads to a more substantial enhancement in the $DP@20$ metric, with a 35.89\% improvement. In comparison, ~\shortname ~ w/o U shows a 20.53\% improvement in $DP@20$.
These results highlight that ~\shortname ~ can be specifically tailored to mitigate one-sided biases and that biases on both user and item sides are interconnected. This emphasizes the importance of simultaneously considering biases on both sides to achieve a more balanced and fair recommendation system.

\subsection{Parameter sensitivity}
In this section, we explore the impact of the hyperparameters $\lambda_1$ and $\lambda_2$ on the outcomes of our recommendation system. We specifically focus on how adjustments to group alignment and global uniformity influence key performance indicators such as accuracy ($NDCG@20$), item-side bias ($PRU$), and user-side bias ($DP@20$). Additionally, we examine the sensitivity of our model to the distribution of entities across the popularity spectrum, from popular to long-tail items.
\begin{figure}[t]
    \centering
    \includegraphics[width=0.9\linewidth]{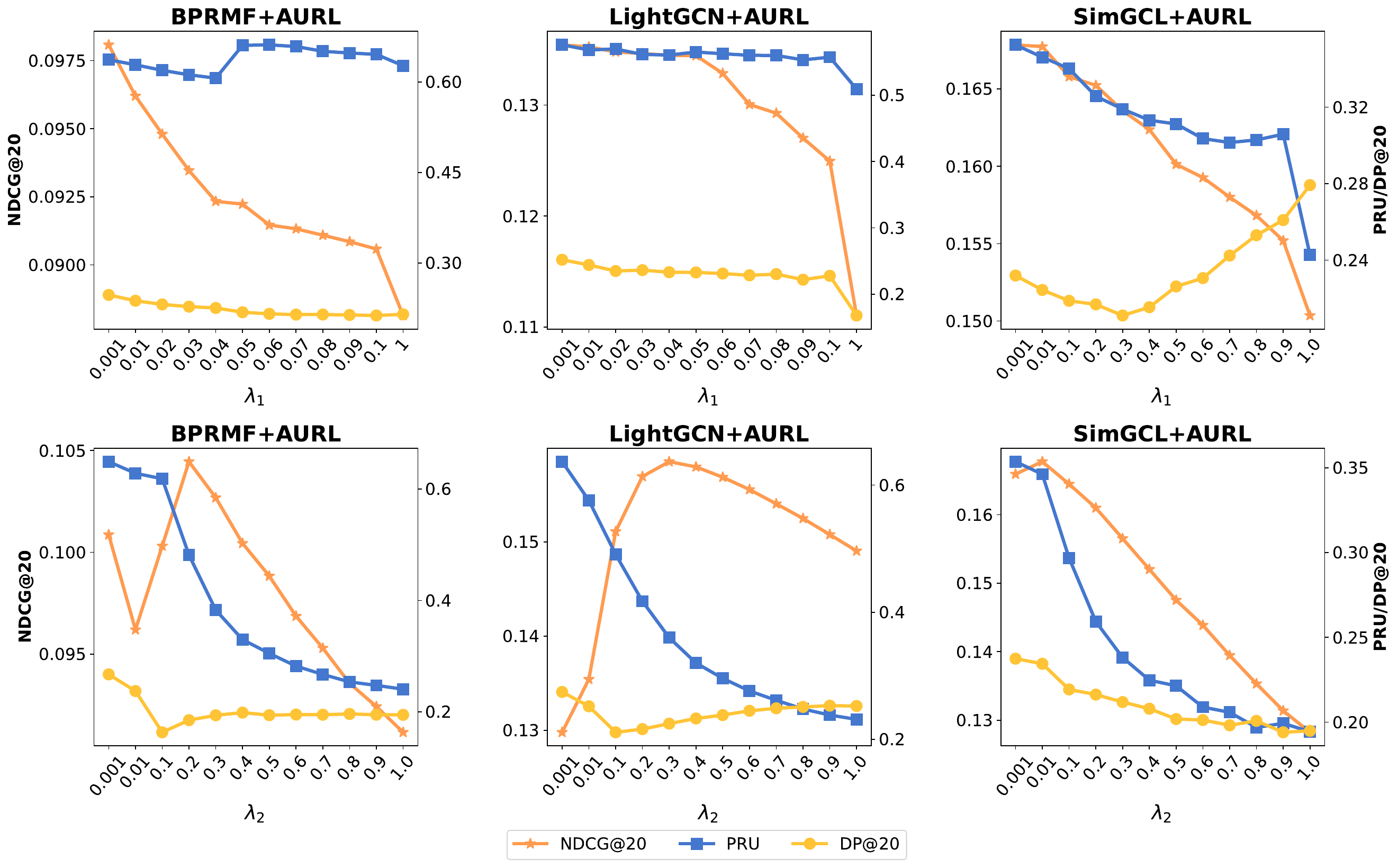}
    \vspace{-0.2cm}
    \caption{Performance comparison w.r.t. different $\lambda_1$ and $\lambda_2$. The top shows the impact of $\lambda_1$ on the results, while the bottom shows the impact of $\lambda_2$ on the results. We present results for three metrics on the Douban-Book dataset: $NDCG@20$, $PRU$, and $DP@20$.}
    \label{fig:Parameter sensitivity}
    \Description{Performance comparison w.r.t. different $\lambda_1$ and $\lambda_2$.}
    \vspace{-0.5cm}
\end{figure}

\subsubsection{\textbf{Impact of the $\lambda_1$.}}
In our objective function, the parameter $\lambda_1$ is crucial for controlling the group-alignment of representation distributions. To assess the effects of varying $\lambda_1$, we systematically modified its value from 0.001 to 1. The results of these variations are presented in the upper three subplots of Fig.~\ref{fig:Parameter sensitivity}.
Our analysis reveals that, across all backbone models, increasing $\lambda_1$ leads to a consistent decline in the $PRU@20$ and $DP@20$ metrics. This pattern underscores the effectiveness of stronger group-alignment constraints in reducing biases within recommendation results. However, it is important to note that there is also a noticeable decrease in the $NDCG@20$ metric, which suggests that while enhancing group-alignment is beneficial for bias mitigation, it may adversely affect the overall recommendation accuracy.


\begin{figure}[t]
    \centering
    \vspace{-0.5cm}
    \subfloat[Item representation distribution]{\includegraphics[width =0.95\linewidth]{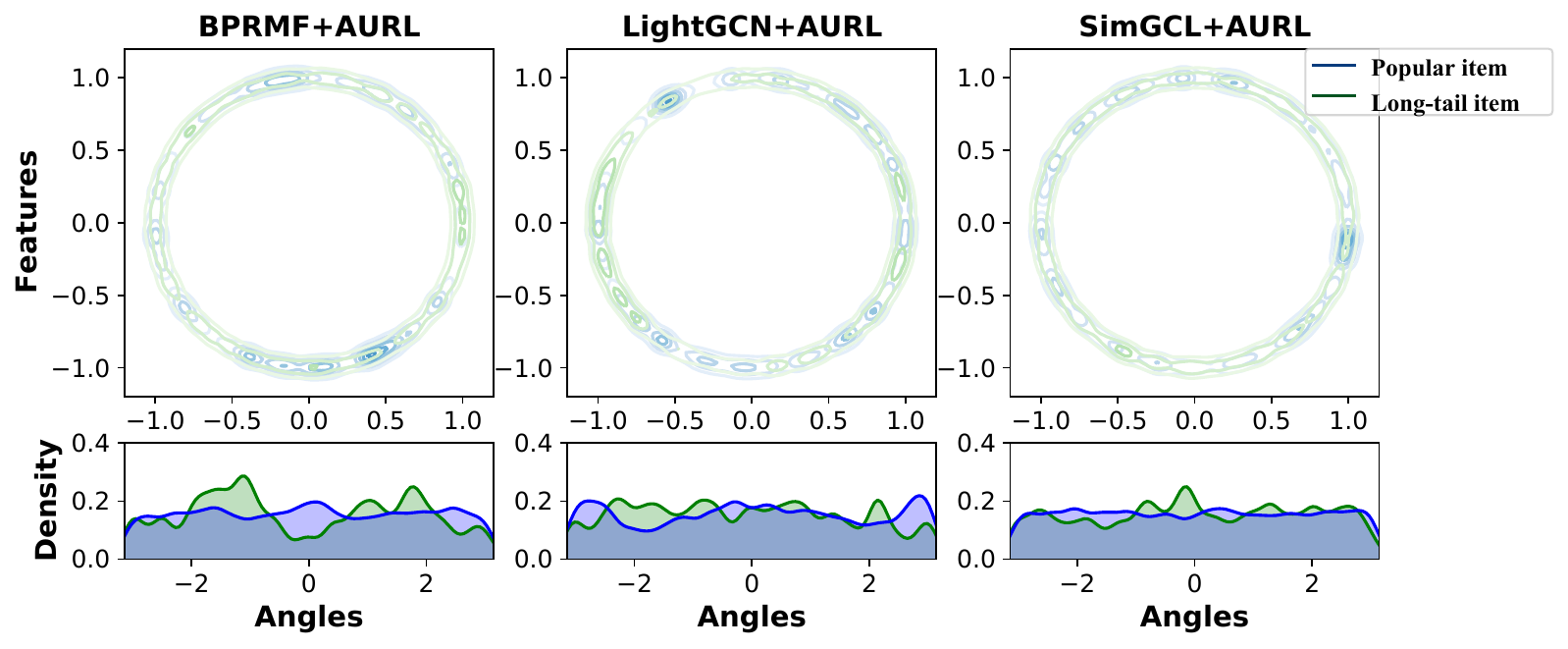}\label{ AU2R Item representation distribution}}
    \hfill
    \subfloat[User representation distribution]{\includegraphics[width =0.95\linewidth]{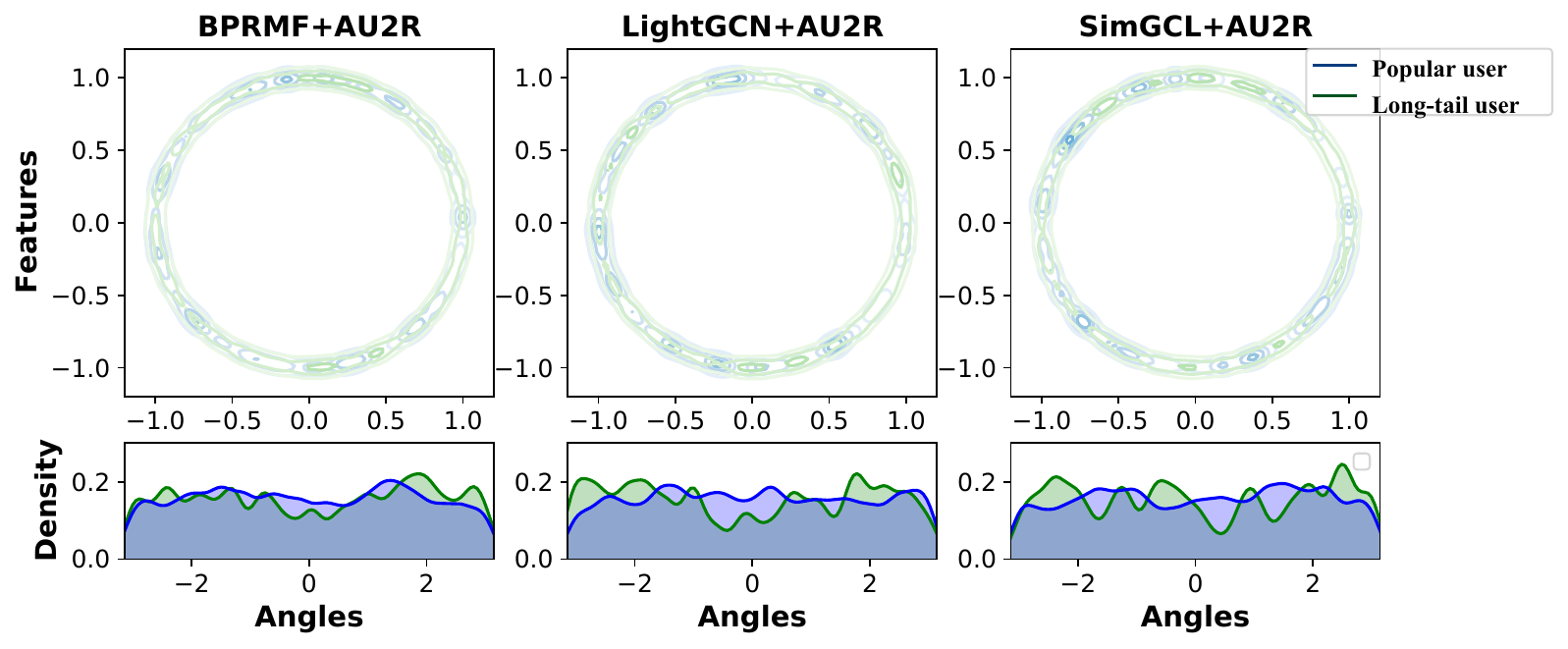}\label{AU2R User representation distribution}}
    \vspace{-0.3cm}
    \caption{Visualization of ~\shortname ~ user and item representations on Douban-Book on $\mathcal{S}^1$. To improve the readability of the figure, we uniformly randomly select 500 items/users for display. We show the entities with Gaussian KDE and the angles with vMF KDE. Green and blue points represent different user or item groups.}
    \label{fig:AU2 Rrepresentation_distribution}
    \Description{Visualization of ~\shortname ~ user and item representations.}
    \vspace{-0.5cm}
\end{figure}

\begin{figure}[t]
    \centering
    \vspace{-0.1cm}
    \includegraphics[width= 0.95\linewidth]{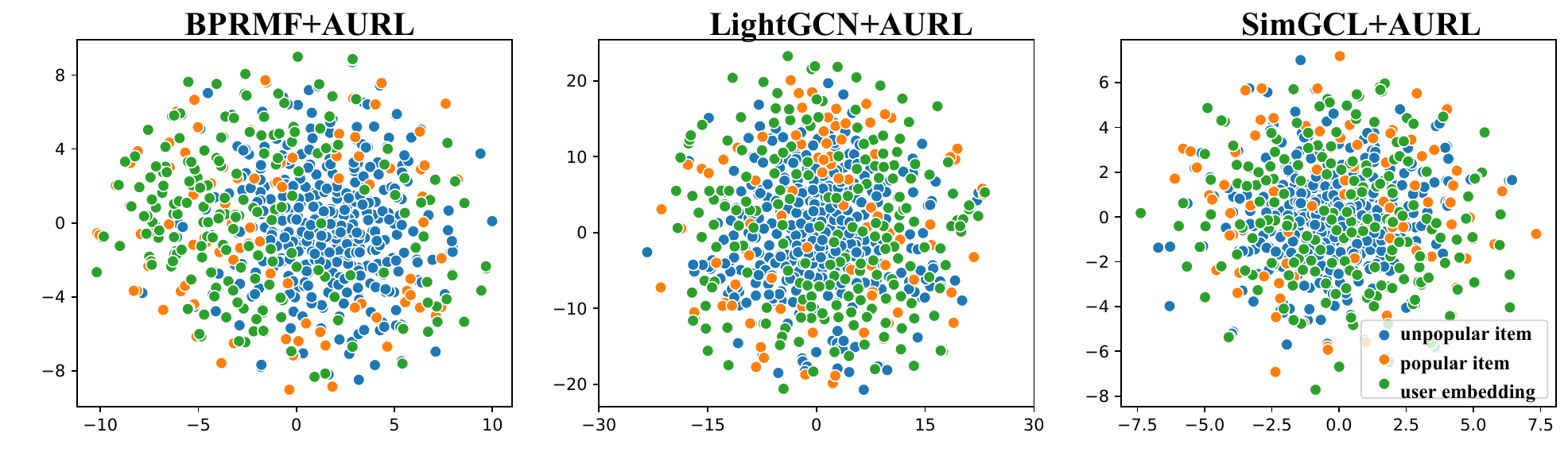}
    \vspace{-0.3cm}
    \caption{Representation visualization of users and items. The green points represent users, and the blue and orange points represent popular items and long-tail items, respectively. To improve the readability of the figure, we uniformly randomly select 500 items/users to show on each subplot after performing t-SNE.}
    \label{fig:IU distirbution}
    \Description{Representation visualization of users and items.}
    \vspace{-0.5cm}
\end{figure}

\subsubsection{\textbf{Impact of the $\lambda_2$.}}
In the lower three subfigures of Fig.~\ref{fig:Parameter sensitivity}, we examine the impact of the global-uniformity hyperparameter, $\lambda_2$. We were pleasantly surprised to discover that an optimal value of $\lambda_2$, particularly around 0.1, not only boosts the accuracy of recommendations but also effectively reduces biases. Specifically, both BPRMF and LightGCN models exhibit peak accuracy at this $\lambda_2$ setting.
Moreover, as we increase the value of $\lambda_2$, the debiasing effect within the models becomes more pronounced. It is important to note, however, that for the SimGCL model, a higher $\lambda_2$ value results in diminished accuracy. This may be attributed to the inherent nature of SimGCL, which incorporates contrastive learning elements that already contribute to a level of global-uniformity. Despite this, our approach continues to effectively mitigate bias within the SimGCL model.

\subsubsection{\textbf{Impact of the division percentage.}}
To assess the sensitivity of our model to the division ratio between popular and long-tail entities, we conducted a series of experiments varying the proportion of entities considered as 'popular'. We defined popular entities as the top 5\%, 10\%, 20\%, and 30\% of entities in the dataset. The outcomes of these experiments are summarized in Table~\ref{zongheng-ccp-host}. The experimental results indicate that, in most scenarios, adjusting the division percentage leads to increased accuracy and reduced bias compared to the best-performing baselines. Specifically, when the proportion of popular users or items is below 10\%, our model tends to show suboptimal results in both accuracy and debiasing effectiveness. This is likely due to the insufficient representation of popular entities, which hampers the model's ability to effectively learn and align their distribution, thus diminishing overall performance. Conversely, increasing the proportion of popular entities generally improves both the accuracy and the debiasing capability of the model, particularly for user-side popularity bias. This enhancement is attributed to the more abundant interaction data available for popular users, which allows the model to more accurately capture their preferences and behavioral patterns. Our findings demonstrate that our model maintains strong performance across various division percentages, highlighting its robustness and adaptability to different entity distribution scenarios.  
\begin{table}
\centering
\refstepcounter{table}
\caption{Performance of AURL across different division percentages of popular users/items. The division percentages represent the proportion of entities considered as popular (top 5\%, top 10\%, top 20\%, and top 30\%) in the Amazon-book and Douban-book datasets.}
\vspace{-0.3cm}
\label{zongheng-ccp-host}
\resizebox{1\columnwidth}{!}{
\begin{tabular}{|c|c|c|c|c|c|c|c|c|c|}
\hline
\multicolumn{2}{|c|}{\textbf{Dataset}}& \multicolumn{4}{c|}{\textbf{Amazon-book}}   & \multicolumn{4}{c|}{\textbf{Douban-book}}    \\
\hline
\textbf{Model}                          & \textbf{pop percentage }&   \textbf{HR$\uparrow$} &\textbf{ NDCG$\uparrow$ }& \textbf{PRU$\downarrow$ }& \textbf{DP$\downarrow$ }       & \textbf{HR$\uparrow$} &\textbf{ NDCG$\uparrow$ }& \textbf{PRU$\downarrow$ }& \textbf{DP$\downarrow$ } \\
\hline
\multirow{6}{*}{\textbf{BPRMF}}  & +Zerosum& 0.0287& 0.0208& 0.4792& 0.2373& 0.1115& 0.0905& 0.6436&0.2029\\
&+InvCF&0.0316 & 0.0230 & 0.5241 & 0.2873&  0.1284 & 0.1016  & 0.6759 & 0.2955\\
 & AURL-5\%          &0.0300&0.0224&    \textbf{0.3497  }  &0.2871& 0.1231 & 0.1043 & 0.4800   & 0.2317  \\
                               & AURL-10\%           & \textbf{0.0327} &\textbf{ 0.0243} & 0.3915 & 0.2918 & 0.1231 & 0.1041 & \textbf{0.4735} & 0.2005  \\
                               &AURL-20\%           & 0.0322 & 0.0235 & 0.4063 &\textbf{ 0.1858 }& 0.1232 & \textbf{0.1045} & 0.4814 & 0.1850   \\
                               &AURL-30\%&0.0325&0.0231&0.4107&0.2649&0.1229&0.1005&0.5456&0.2128\\
                            
\hline
\multirow{6}{*}{\textbf{LightGCN}}&+Zerosum&0.0376&0.0274&0.4911&0.2736&0.1534&   0.1256  &   0.5719    &0.2536 \\
&+InvCF&0.0409&0.0309&0.5028&0.2691&0.1690&0.1388&0.6260&0.2656\\
& AURL-5\%          & 0.0459       &  \textbf{ 0.0350 }   & 0.3661      &     \textbf{0.2398 }  & 0.1830  & 0.1584 & 0.3579 & 0.2900    \\
                               & AURL-10\%           &\textbf{ 0.0460}  & 0.0349 & 0.3633 & 0.2410  & 0.1829 & 0.1583 & \textbf{0.3577} & 0.2452  \\
                               & AURL-20\%           & 0.0458 & 0.0349 & \textbf{0.3543} & 0.2399 &\textbf{ 0.1831} & \textbf{0.1585 }& 0.3599 & \textbf{0.2247 } \\
                                &AURL-30\%&0.0451&0.0340&0.3891&0.2525&0.1807&0.1544&0.3648&0.2329\\
\hline
\multirow{6}{*}{\textbf{SimGCL}} &+Zerosum&0.0506&0.0403&0.2136&\underline{0.2438}& 0.1865&    0.1570 &  0.3450    &0.2378\\
&+InvCF&0.0540&0.0411&0.2486&0.2847&0.1846&0.1539&0.3847&0.2730\\
& AURL-5\%          &0.0491&0.0374&0.3138&0.2565& 0.1845 & 0.1611 & \textbf{0.2712 }& 0.2809  \\
                               & AURL-10\%           & 0.0502 & 0.0380  & 0.2903 & 0.2677 & 0.1879 & 0.1641 & 0.2828 & 0.2424  \\
                               & AURL-20\%           & 0.0541 & 0.0418 & \textbf{0.1875} & \textbf{0.2281} & 0.1884 & 0.1644 & 0.2967 & \textbf{0.2192 }\\
                                &AURL-30\%&0.0537&0.0414&0.2133&0.2485&0.1893&0.1656&0.2999&0.2241\\
\hline
\end{tabular}}
\vspace{-0.5cm}
\end{table}

\subsection{Visualizing the distribution of representations}
To intuitively demonstrate how ~\shortname ~ mitigates recommendation biases, we employed t-SNE~\cite{van2008visualizing} to visualize user and item representations within the Douban-Book dataset. The distributions of representations under various backbones facilitated by ~\shortname ~ are depicted in Fig.~\ref{fig:AU2 Rrepresentation_distribution} and ~\ref{fig:IU distirbution}.

Compared to the baseline shown in Fig.~\ref{fig:representation_distribution}, the representations learned through our method display a more even and consistent distribution across the space. Notably, both popular and long-tail entity groups are uniformly dispersed and become indistinguishable from one another, enhancing the likelihood of achieving unbiased recommendation results. This uniform distribution indicates that ~\shortname ~ effectively mitigates biases on both the user and item sides as intended.
Further comparisons with Fig.~\ref{fig:fc} reveal that in the representation space, users are more uniformly dispersed and are not predominantly clustered around popular items. This alteration suggests that the model recommendations are now more heavily influenced by genuine user preferences and item characteristics rather than by the popularity bias. Consequently, both popular and long-tail items are less distinguishable, leading to a more consistent distribution across the board.

Moreover, we discuss the inherent trade-off between effectiveness and debiasing. By constraining the representation distribution, ~\shortname ~ modifies the natural grouping of users and items, forcing a greater similarity between different groups. While this approach enhances debiasing, it may occasionally reduce the effectiveness of recommendations due to altered distribution dynamics. 

\section{Conclusion and future work}

In this paper, we analyzed deep-rooted biases in recommender systems, focusing on group-discrepancy and global-collapse in representation distribution, which led to biased outcomes. We introduced a framework, \textit{\shortname}, that addressed these biases by enforcing group-alignment and global-uniformity. This approach was compatible with any CF-based model as an auxiliary task, requiring no extra parameter tuning. Our extensive evaluations across multiple domains using three datasets demonstrated that \textit{\shortname} outperformed existing baselines in debiasing, effectively reducing user and item biases while maintaining accuracy.
For future work, we aim to identify and mitigate additional types of biases and expand our model's application to broader scenarios, thereby enhancing the fairness and accuracy of recommender systems.

\begin{acks}
This work was supported in part by grants from the National Key Research and Development Program of China (Grant No. 2021ZD0111802), the National Natural Science Foundation of China (Grant Nos. U23B2031, 72188101), the China Postdoctoral Science Foundation (Grant No. 2023M741943), and the Postdoctoral Fellowship Program of CPSF (Grant No. GZC20231373).
\end{acks}

\bibliographystyle{ACM-Reference-Format}
\bibliography{sample-base}


\begin{thebibliography}{66}


\ifx \showCODEN    \undefined \def \showCODEN     #1{\unskip}     \fi
\ifx \showDOI      \undefined \def \showDOI       #1{#1}\fi
\ifx \showISBNx    \undefined \def \showISBNx     #1{\unskip}     \fi
\ifx \showISBNxiii \undefined \def \showISBNxiii  #1{\unskip}     \fi
\ifx \showISSN     \undefined \def \showISSN      #1{\unskip}     \fi
\ifx \showLCCN     \undefined \def \showLCCN      #1{\unskip}     \fi
\ifx \shownote     \undefined \def \shownote      #1{#1}          \fi
\ifx \showarticletitle \undefined \def \showarticletitle #1{#1}   \fi
\ifx \showURL      \undefined \def \showURL       {\relax}        \fi
\providecommand\bibfield[2]{#2}
\providecommand\bibinfo[2]{#2}
\providecommand\natexlab[1]{#1}
\providecommand\showeprint[2][]{arXiv:#2}

\bibitem[Amari(1993)]%
        {amari1993backpropagation}
\bibfield{author}{\bibinfo{person}{Shun-ichi Amari}.} \bibinfo{year}{1993}\natexlab{}.
\newblock \showarticletitle{Backpropagation and stochastic gradient descent method}.
\newblock \bibinfo{journal}{\emph{Neurocomputing}} (\bibinfo{year}{1993}).
\newblock


\bibitem[Bonner and Vasile(2018)]%
        {bonner2018causal}
\bibfield{author}{\bibinfo{person}{Stephen Bonner} {and} \bibinfo{person}{Flavian Vasile}.} \bibinfo{year}{2018}\natexlab{}.
\newblock \showarticletitle{Causal embeddings for recommendation}.
\newblock \bibinfo{journal}{\emph{RecSys}} (\bibinfo{year}{2018}).
\newblock


\bibitem[Borodachov et~al\mbox{.}(2019)]%
        {borodachov2019discrete}
\bibfield{author}{\bibinfo{person}{Sergiy~V Borodachov}, \bibinfo{person}{Douglas~P Hardin}, {and} \bibinfo{person}{Edward~B Saff}.} \bibinfo{year}{2019}\natexlab{}.
\newblock \showarticletitle{Discrete energy on rectifiable sets}.
\newblock \bibinfo{journal}{\emph{Springer}} (\bibinfo{year}{2019}).
\newblock


\bibitem[Chen et~al\mbox{.}(2019)]%
        {chen2019efficient}
\bibfield{author}{\bibinfo{person}{Chong Chen}, \bibinfo{person}{Min Zhang}, \bibinfo{person}{Chenyang Wang}, \bibinfo{person}{Weizhi Ma}, \bibinfo{person}{Minming Li}, \bibinfo{person}{Yiqun Liu}, {and} \bibinfo{person}{Shaoping Ma}.} \bibinfo{year}{2019}\natexlab{}.
\newblock \showarticletitle{An efficient adaptive transfer neural network for social-aware recommendation}.
\newblock \bibinfo{journal}{\emph{SIGIR}} (\bibinfo{year}{2019}).
\newblock


\bibitem[Chen et~al\mbox{.}(2021)]%
        {AutoDebias}
\bibfield{author}{\bibinfo{person}{Jiawei Chen}, \bibinfo{person}{Hande Dong}, \bibinfo{person}{Yang Qiu}, \bibinfo{person}{Xiangnan He}, \bibinfo{person}{Xin Xin}, \bibinfo{person}{Liang Chen}, \bibinfo{person}{Guli Lin}, {and} \bibinfo{person}{Keping Yang}.} \bibinfo{year}{2021}\natexlab{}.
\newblock \showarticletitle{AutoDebias: Learning to Debias for Recommendation}.
\newblock \bibinfo{journal}{\emph{SIGIR}} (\bibinfo{year}{2021}).
\newblock


\bibitem[Chen et~al\mbox{.}(2020a)]%
        {Chen2020BiasAD}
\bibfield{author}{\bibinfo{person}{Jiawei Chen}, \bibinfo{person}{Hande Dong}, \bibinfo{person}{Xiang Wang}, \bibinfo{person}{Fuli Feng}, \bibinfo{person}{Meng Wang}, {and} \bibinfo{person}{Xiangnan He}.} \bibinfo{year}{2020}\natexlab{a}.
\newblock \showarticletitle{Bias and Debias in Recommender System: A Survey and Future Directions}.
\newblock \bibinfo{journal}{\emph{TOIS}} (\bibinfo{year}{2020}).
\newblock


\bibitem[Chen et~al\mbox{.}(2024)]%
        {chen2024graph}
\bibfield{author}{\bibinfo{person}{Jiajia Chen}, \bibinfo{person}{Jiancan Wu}, \bibinfo{person}{Jiawei Chen}, \bibinfo{person}{Xin Xin}, \bibinfo{person}{Yong Li}, {and} \bibinfo{person}{Xiangnan He}.} \bibinfo{year}{2024}\natexlab{}.
\newblock \showarticletitle{How graph convolutions amplify popularity bias for recommendation?}
\newblock \bibinfo{journal}{\emph{FCS}} (\bibinfo{year}{2024}).
\newblock


\bibitem[Chen et~al\mbox{.}(2017)]%
        {attchen2017attentive}
\bibfield{author}{\bibinfo{person}{Jingyuan Chen}, \bibinfo{person}{Hanwang Zhang}, \bibinfo{person}{Xiangnan He}, \bibinfo{person}{Liqiang Nie}, \bibinfo{person}{Wei Liu}, {and} \bibinfo{person}{Tat-Seng Chua}.} \bibinfo{year}{2017}\natexlab{}.
\newblock \showarticletitle{Attentive collaborative filtering: Multimedia recommendation with item-and component-level attention}.
\newblock \bibinfo{journal}{\emph{SIGIR}} (\bibinfo{year}{2017}).
\newblock


\bibitem[Chen et~al\mbox{.}(2020b)]%
        {Chen2020RevisitingGB}
\bibfield{author}{\bibinfo{person}{Lei Chen}, \bibinfo{person}{Le Wu}, \bibinfo{person}{Richang Hong}, \bibinfo{person}{Kun Zhang}, {and} \bibinfo{person}{Meng Wang}.} \bibinfo{year}{2020}\natexlab{b}.
\newblock \showarticletitle{Revisiting Graph based Collaborative Filtering: A Linear Residual Graph Convolutional Network Approach}.
\newblock \bibinfo{journal}{\emph{AAAI}} (\bibinfo{year}{2020}).
\newblock


\bibitem[Chen et~al\mbox{.}(2023)]%
        {fairdata23}
\bibfield{author}{\bibinfo{person}{Lei Chen}, \bibinfo{person}{Le Wu}, \bibinfo{person}{Kun Zhang}, \bibinfo{person}{Richang Hong}, \bibinfo{person}{Defu Lian}, \bibinfo{person}{Zhiqiang Zhang}, \bibinfo{person}{Jun Zhou}, {and} \bibinfo{person}{Meng Wang}.} \bibinfo{year}{2023}\natexlab{}.
\newblock \showarticletitle{Improving Recommendation Fairness via Data Augmentation}.
\newblock \bibinfo{journal}{\emph{WWW}} (\bibinfo{year}{2023}).
\newblock


\bibitem[Cui et~al\mbox{.}(2012)]%
        {cui2012discover}
\bibfield{author}{\bibinfo{person}{Anqi Cui}, \bibinfo{person}{Min Zhang}, \bibinfo{person}{Yiqun Liu}, \bibinfo{person}{Shaoping Ma}, {and} \bibinfo{person}{Kuo Zhang}.} \bibinfo{year}{2012}\natexlab{}.
\newblock \showarticletitle{Discover breaking events with popular hashtags in twitter}.
\newblock \bibinfo{journal}{\emph{ICKM}} (\bibinfo{year}{2012}).
\newblock


\bibitem[Deng et~al\mbox{.}(2022)]%
        {deng2022graph}
\bibfield{author}{\bibinfo{person}{Leyan Deng}, \bibinfo{person}{Defu Lian}, \bibinfo{person}{Chenwang Wu}, {and} \bibinfo{person}{Enhong Chen}.} \bibinfo{year}{2022}\natexlab{}.
\newblock \showarticletitle{Graph convolution network based recommender systems: Learning guarantee and item mixture powered strategy}.
\newblock \bibinfo{journal}{\emph{NeurIPS}} (\bibinfo{year}{2022}).
\newblock


\bibitem[Gao et~al\mbox{.}(2023a)]%
        {gao2023alleviating}
\bibfield{author}{\bibinfo{person}{Chongming Gao}, \bibinfo{person}{Kexin Huang}, \bibinfo{person}{Jiawei Chen}, \bibinfo{person}{Yuan Zhang}, \bibinfo{person}{Biao Li}, \bibinfo{person}{Peng Jiang}, \bibinfo{person}{Shiqi Wang}, \bibinfo{person}{Zhong Zhang}, {and} \bibinfo{person}{Xiangnan He}.} \bibinfo{year}{2023}\natexlab{a}.
\newblock \showarticletitle{Alleviating matthew effect of offline reinforcement learning in interactive recommendation}.
\newblock \bibinfo{journal}{\emph{SIGIR}} (\bibinfo{year}{2023}).
\newblock


\bibitem[Gao et~al\mbox{.}(2023b)]%
        {gao2023cirs}
\bibfield{author}{\bibinfo{person}{Chongming Gao}, \bibinfo{person}{Shiqi Wang}, \bibinfo{person}{Shijun Li}, \bibinfo{person}{Jiawei Chen}, \bibinfo{person}{Xiangnan He}, \bibinfo{person}{Wenqiang Lei}, \bibinfo{person}{Biao Li}, \bibinfo{person}{Yuan Zhang}, {and} \bibinfo{person}{Peng Jiang}.} \bibinfo{year}{2023}\natexlab{b}.
\newblock \showarticletitle{CIRS: Bursting filter bubbles by counterfactual interactive recommender system}.
\newblock \bibinfo{journal}{\emph{TOIS}} (\bibinfo{year}{2023}).
\newblock


\bibitem[Glorot and Bengio(2010)]%
        {glorot2010understanding}
\bibfield{author}{\bibinfo{person}{Xavier Glorot} {and} \bibinfo{person}{Yoshua Bengio}.} \bibinfo{year}{2010}\natexlab{}.
\newblock \showarticletitle{Understanding the difficulty of training deep feedforward neural networks}.
\newblock \bibinfo{journal}{\emph{AISTATS}} (\bibinfo{year}{2010}).
\newblock


\bibitem[Gretton et~al\mbox{.}(2012)]%
        {gretton2012kernel}
\bibfield{author}{\bibinfo{person}{Arthur Gretton}, \bibinfo{person}{Karsten~M Borgwardt}, \bibinfo{person}{Malte~J Rasch}, \bibinfo{person}{Bernhard Sch{\"o}lkopf}, {and} \bibinfo{person}{Alexander Smola}.} \bibinfo{year}{2012}\natexlab{}.
\newblock \showarticletitle{A kernel two-sample test}.
\newblock \bibinfo{journal}{\emph{JMLR}} (\bibinfo{year}{2012}).
\newblock


\bibitem[Harper and Konstan(2015)]%
        {harper2015movielens}
\bibfield{author}{\bibinfo{person}{F~Maxwell Harper} {and} \bibinfo{person}{Joseph~A Konstan}.} \bibinfo{year}{2015}\natexlab{}.
\newblock \showarticletitle{The movielens datasets: History and context}.
\newblock \bibinfo{journal}{\emph{TIIS}} (\bibinfo{year}{2015}).
\newblock


\bibitem[He et~al\mbox{.}(2020)]%
        {He2020LightGCNSA}
\bibfield{author}{\bibinfo{person}{Xiangnan He}, \bibinfo{person}{Kuan Deng}, \bibinfo{person}{Xiang Wang}, \bibinfo{person}{Yan Li}, \bibinfo{person}{Yongdong Zhang}, {and} \bibinfo{person}{Meng Wang}.} \bibinfo{year}{2020}\natexlab{}.
\newblock \showarticletitle{LightGCN: Simplifying and Powering Graph Convolution Network for Recommendation}.
\newblock \bibinfo{journal}{\emph{SIGIR}} (\bibinfo{year}{2020}).
\newblock


\bibitem[He et~al\mbox{.}(2017)]%
        {WWW2017NCF}
\bibfield{author}{\bibinfo{person}{Xiangnan He}, \bibinfo{person}{Lizi Liao}, \bibinfo{person}{Hanwang Zhang}, \bibinfo{person}{Liqiang Nie}, \bibinfo{person}{Xia Hu}, {and} \bibinfo{person}{Tat-Seng Chua}.} \bibinfo{year}{2017}\natexlab{}.
\newblock \showarticletitle{Neural collaborative filtering}.
\newblock \bibinfo{journal}{\emph{WWW}} (\bibinfo{year}{2017}).
\newblock


\bibitem[Hou et~al\mbox{.}(2019)]%
        {hou2019explainable}
\bibfield{author}{\bibinfo{person}{Min Hou}, \bibinfo{person}{Le Wu}, \bibinfo{person}{Enhong Chen}, \bibinfo{person}{Zhi Li}, \bibinfo{person}{Vincent~W Zheng}, {and} \bibinfo{person}{Qi Liu}.} \bibinfo{year}{2019}\natexlab{}.
\newblock \showarticletitle{Explainable fashion recommendation: A semantic attribute region guided approach}.
\newblock \bibinfo{journal}{\emph{IJCAI}} (\bibinfo{year}{2019}).
\newblock


\bibitem[Kiefer and Wolfowitz(1952)]%
        {kiefer1952stochastic}
\bibfield{author}{\bibinfo{person}{Jack Kiefer} {and} \bibinfo{person}{Jacob Wolfowitz}.} \bibinfo{year}{1952}\natexlab{}.
\newblock \showarticletitle{Stochastic estimation of the maximum of a regression function}.
\newblock \bibinfo{journal}{\emph{JSTOR}} (\bibinfo{year}{1952}).
\newblock


\bibitem[Kingma and Ba(2015)]%
        {DBLP:journals/corr/KingmaB14}
\bibfield{author}{\bibinfo{person}{Diederik~P. Kingma} {and} \bibinfo{person}{Jimmy Ba}.} \bibinfo{year}{2015}\natexlab{}.
\newblock \showarticletitle{Adam: {A} Method for Stochastic Optimization}.
\newblock \bibinfo{journal}{\emph{ICLR}} (\bibinfo{year}{2015}).
\newblock


\bibitem[Koren et~al\mbox{.}(2009)]%
        {Koren2009MatrixFT}
\bibfield{author}{\bibinfo{person}{Yehuda Koren}, \bibinfo{person}{Robert~M. Bell}, {and} \bibinfo{person}{Chris Volinsky}.} \bibinfo{year}{2009}\natexlab{}.
\newblock \showarticletitle{Matrix Factorization Techniques for Recommender Systems}.
\newblock \bibinfo{journal}{\emph{Computer}} (\bibinfo{year}{2009}).
\newblock


\bibitem[Kusner et~al\mbox{.}(2017)]%
        {kusner2017counterfactual}
\bibfield{author}{\bibinfo{person}{Matt~J Kusner}, \bibinfo{person}{Joshua Loftus}, \bibinfo{person}{Chris Russell}, {and} \bibinfo{person}{Ricardo Silva}.} \bibinfo{year}{2017}\natexlab{}.
\newblock \showarticletitle{Counterfactual fairness}.
\newblock \bibinfo{journal}{\emph{NeurIPS}} (\bibinfo{year}{2017}).
\newblock


\bibitem[Li et~al\mbox{.}(2021)]%
        {li2021user}
\bibfield{author}{\bibinfo{person}{Yunqi Li}, \bibinfo{person}{Hanxiong Chen}, \bibinfo{person}{Zuohui Fu}, \bibinfo{person}{Yingqiang Ge}, {and} \bibinfo{person}{Yongfeng Zhang}.} \bibinfo{year}{2021}\natexlab{}.
\newblock \showarticletitle{User-oriented fairness in recommendation}.
\newblock \bibinfo{journal}{\emph{WWW}} (\bibinfo{year}{2021}).
\newblock


\bibitem[Liu et~al\mbox{.}(2023)]%
        {liu2023mitigating}
\bibfield{author}{\bibinfo{person}{Zhongzhou Liu}, \bibinfo{person}{Yuan Fang}, {and} \bibinfo{person}{Min Wu}.} \bibinfo{year}{2023}\natexlab{}.
\newblock \showarticletitle{Mitigating popularity bias for users and items with fairness-centric adaptive recommendation}.
\newblock \bibinfo{journal}{\emph{TOIS}} (\bibinfo{year}{2023}).
\newblock


\bibitem[Naghiaei et~al\mbox{.}(2022)]%
        {naghiaei2022cpfair}
\bibfield{author}{\bibinfo{person}{Mohammadmehdi Naghiaei}, \bibinfo{person}{Hossein~A Rahmani}, {and} \bibinfo{person}{Yashar Deldjoo}.} \bibinfo{year}{2022}\natexlab{}.
\newblock \showarticletitle{Cpfair: Personalized consumer and producer fairness re-ranking for recommender systems}.
\newblock \bibinfo{journal}{\emph{SIGIR}} (\bibinfo{year}{2022}).
\newblock


\bibitem[Perc(2014)]%
        {perc2014matthew}
\bibfield{author}{\bibinfo{person}{Matja{\v{z}} Perc}.} \bibinfo{year}{2014}\natexlab{}.
\newblock \showarticletitle{The Matthew effect in empirical data}.
\newblock \bibinfo{journal}{\emph{J R Soc Interface}} (\bibinfo{year}{2014}).
\newblock


\bibitem[Rahman et~al\mbox{.}(2020)]%
        {rahman2020minimum}
\bibfield{author}{\bibinfo{person}{Mohammad~Mahfujur Rahman}, \bibinfo{person}{Clinton Fookes}, \bibinfo{person}{Mahsa Baktashmotlagh}, {and} \bibinfo{person}{Sridha Sridharan}.} \bibinfo{year}{2020}\natexlab{}.
\newblock \showarticletitle{On minimum discrepancy estimation for deep domain adaptation}.
\newblock \bibinfo{journal}{\emph{DAVU}} (\bibinfo{year}{2020}).
\newblock


\bibitem[Ren et~al\mbox{.}(2022b)]%
        {ren2022mitigating}
\bibfield{author}{\bibinfo{person}{Weijieying Ren}, \bibinfo{person}{Lei Wang}, \bibinfo{person}{Kunpeng Liu}, \bibinfo{person}{Ruocheng Guo}, \bibinfo{person}{Lim~Ee Peng}, {and} \bibinfo{person}{Yanjie Fu}.} \bibinfo{year}{2022}\natexlab{b}.
\newblock \showarticletitle{Mitigating popularity bias in recommendation with unbalanced interactions: A gradient perspective}.
\newblock \bibinfo{journal}{\emph{ICDM}} (\bibinfo{year}{2022}).
\newblock


\bibitem[Ren et~al\mbox{.}(2022a)]%
        {ren2022semi}
\bibfield{author}{\bibinfo{person}{Weijieying Ren}, \bibinfo{person}{Pengyang Wang}, \bibinfo{person}{Xiaolin Li}, \bibinfo{person}{Charles~E Hughes}, {and} \bibinfo{person}{Yanjie Fu}.} \bibinfo{year}{2022}\natexlab{a}.
\newblock \showarticletitle{Semi-supervised drifted stream learning with short lookback}.
\newblock \bibinfo{journal}{\emph{KDD}} (\bibinfo{year}{2022}).
\newblock


\bibitem[Rendle and Freudenthaler(2014)]%
        {rendle2014improving}
\bibfield{author}{\bibinfo{person}{Steffen Rendle} {and} \bibinfo{person}{Christoph Freudenthaler}.} \bibinfo{year}{2014}\natexlab{}.
\newblock \showarticletitle{Improving pairwise learning for item recommendation from implicit feedback}.
\newblock \bibinfo{journal}{\emph{WSDM}} (\bibinfo{year}{2014}).
\newblock


\bibitem[Rendle et~al\mbox{.}(2009)]%
        {Rendle2009BPRBP}
\bibfield{author}{\bibinfo{person}{Steffen Rendle}, \bibinfo{person}{Christoph Freudenthaler}, \bibinfo{person}{Zeno Gantner}, {and} \bibinfo{person}{Lars Schmidt-Thieme}.} \bibinfo{year}{2009}\natexlab{}.
\newblock \showarticletitle{BPR: Bayesian Personalized Ranking from Implicit Feedback}.
\newblock \bibinfo{journal}{\emph{UAI}} (\bibinfo{year}{2009}).
\newblock


\bibitem[Rhee et~al\mbox{.}(2022)]%
        {rhee2022countering}
\bibfield{author}{\bibinfo{person}{Wondo Rhee}, \bibinfo{person}{Sung~Min Cho}, {and} \bibinfo{person}{Bongwon Suh}.} \bibinfo{year}{2022}\natexlab{}.
\newblock \showarticletitle{Countering Popularity Bias by Regularizing Score Differences}.
\newblock \bibinfo{journal}{\emph{RecSys}} (\bibinfo{year}{2022}).
\newblock


\bibitem[Sebastian(2023)]%
        {sebastian2023malayalam}
\bibfield{author}{\bibinfo{person}{Mary~Priya Sebastian}.} \bibinfo{year}{2023}\natexlab{}.
\newblock \showarticletitle{Malayalam Natural Language Processing: Challenges in Building a Phrase-Based Statistical Machine Translation System}.
\newblock \bibinfo{journal}{\emph{TALLIP}} (\bibinfo{year}{2023}).
\newblock


\bibitem[Shao et~al\mbox{.}(2022)]%
        {shao2022faircf}
\bibfield{author}{\bibinfo{person}{Pengyang Shao}, \bibinfo{person}{Le Wu}, \bibinfo{person}{Lei Chen}, \bibinfo{person}{Kun Zhang}, {and} \bibinfo{person}{Meng Wang}.} \bibinfo{year}{2022}\natexlab{}.
\newblock \showarticletitle{FairCF: Fairness-aware collaborative filtering}.
\newblock \bibinfo{journal}{\emph{SCIS}} (\bibinfo{year}{2022}).
\newblock


\bibitem[Shen et~al\mbox{.}(2017)]%
        {shen2017style}
\bibfield{author}{\bibinfo{person}{Tianxiao Shen}, \bibinfo{person}{Tao Lei}, \bibinfo{person}{Regina Barzilay}, {and} \bibinfo{person}{Tommi Jaakkola}.} \bibinfo{year}{2017}\natexlab{}.
\newblock \showarticletitle{Style transfer from non-parallel text by cross-alignment}.
\newblock \bibinfo{journal}{\emph{NeurIPS}} (\bibinfo{year}{2017}).
\newblock


\bibitem[Tolstikhin et~al\mbox{.}(2016)]%
        {Tolstikhin2016MinimaxEO}
\bibfield{author}{\bibinfo{person}{Ilya~O. Tolstikhin}, \bibinfo{person}{Bharath~K. Sriperumbudur}, {and} \bibinfo{person}{Bernhard Sch{\"o}lkopf}.} \bibinfo{year}{2016}\natexlab{}.
\newblock \showarticletitle{Minimax Estimation of Maximum Mean Discrepancy with Radial Kernels}.
\newblock \bibinfo{journal}{\emph{NeurIPS}} (\bibinfo{year}{2016}).
\newblock


\bibitem[Van~der Maaten and Hinton(2008)]%
        {van2008visualizing}
\bibfield{author}{\bibinfo{person}{Laurens Van~der Maaten} {and} \bibinfo{person}{Geoffrey Hinton}.} \bibinfo{year}{2008}\natexlab{}.
\newblock \showarticletitle{Visualizing data using t-SNE.}
\newblock \bibinfo{journal}{\emph{J Mach Learn Res}} (\bibinfo{year}{2008}).
\newblock


\bibitem[Viehmann(2021)]%
        {viehmann2021partial}
\bibfield{author}{\bibinfo{person}{Thomas Viehmann}.} \bibinfo{year}{2021}\natexlab{}.
\newblock \showarticletitle{Partial Wasserstein and Maximum Mean Discrepancy distances for bridging the gap between outlier detection and drift detection}.
\newblock \bibinfo{journal}{\emph{arXiv}} (\bibinfo{year}{2021}).
\newblock


\bibitem[Wang et~al\mbox{.}(2022b)]%
        {Wang2022TowardsRA}
\bibfield{author}{\bibinfo{person}{Chenyang Wang}, \bibinfo{person}{Yuanqing Yu}, \bibinfo{person}{Weizhi Ma}, \bibinfo{person}{M. Zhang}, \bibinfo{person}{C. Chen}, \bibinfo{person}{Yiqun Liu}, {and} \bibinfo{person}{Shaoping Ma}.} \bibinfo{year}{2022}\natexlab{b}.
\newblock \showarticletitle{Towards Representation Alignment and Uniformity in Collaborative Filtering}.
\newblock \bibinfo{journal}{\emph{KDD}} (\bibinfo{year}{2022}).
\newblock


\bibitem[Wang and Isola(2020)]%
        {wang2020understanding}
\bibfield{author}{\bibinfo{person}{Tongzhou Wang} {and} \bibinfo{person}{Phillip Isola}.} \bibinfo{year}{2020}\natexlab{}.
\newblock \showarticletitle{Understanding contrastive representation learning through alignment and uniformity on the hypersphere}.
\newblock \bibinfo{journal}{\emph{ICML}} (\bibinfo{year}{2020}).
\newblock


\bibitem[Wang et~al\mbox{.}(2021)]%
        {DecRS}
\bibfield{author}{\bibinfo{person}{Wenjie Wang}, \bibinfo{person}{Fuli Feng}, \bibinfo{person}{Xiangnan He}, \bibinfo{person}{Xiang Wang}, {and} \bibinfo{person}{Tat-Seng Chua}.} \bibinfo{year}{2021}\natexlab{}.
\newblock \showarticletitle{Deconfounded Recommendation for Alleviating Bias Amplification}.
\newblock \bibinfo{journal}{\emph{KDD}} (\bibinfo{year}{2021}).
\newblock


\bibitem[Wang et~al\mbox{.}(2022a)]%
        {wang2022causal}
\bibfield{author}{\bibinfo{person}{Wenjie Wang}, \bibinfo{person}{Xinyu Lin}, \bibinfo{person}{Fuli Feng}, \bibinfo{person}{Xiangnan He}, \bibinfo{person}{Min Lin}, {and} \bibinfo{person}{Tat-Seng Chua}.} \bibinfo{year}{2022}\natexlab{a}.
\newblock \showarticletitle{Causal representation learning for out-of-distribution recommendation}.
\newblock \bibinfo{journal}{\emph{WWW}} (\bibinfo{year}{2022}).
\newblock


\bibitem[Wang et~al\mbox{.}(2019)]%
        {Wang2019NeuralGC}
\bibfield{author}{\bibinfo{person}{Xiang Wang}, \bibinfo{person}{Xiangnan He}, \bibinfo{person}{Meng Wang}, \bibinfo{person}{Fuli Feng}, {and} \bibinfo{person}{Tat-Seng Chua}.} \bibinfo{year}{2019}\natexlab{}.
\newblock \showarticletitle{Neural Graph Collaborative Filtering}.
\newblock \bibinfo{journal}{\emph{SIGIR}} (\bibinfo{year}{2019}).
\newblock


\bibitem[Wang et~al\mbox{.}(2023)]%
        {wang2023survey}
\bibfield{author}{\bibinfo{person}{Yifan Wang}, \bibinfo{person}{Weizhi Ma}, \bibinfo{person}{Min Zhang}, \bibinfo{person}{Yiqun Liu}, {and} \bibinfo{person}{Shaoping Ma}.} \bibinfo{year}{2023}\natexlab{}.
\newblock \showarticletitle{A survey on the fairness of recommender systems}.
\newblock \bibinfo{journal}{\emph{TOIS}} (\bibinfo{year}{2023}).
\newblock


\bibitem[Wei et~al\mbox{.}(2020)]%
        {Wei2020ModelAgnosticCR}
\bibfield{author}{\bibinfo{person}{Tianxin Wei}, \bibinfo{person}{Fuli Feng}, \bibinfo{person}{Jiawei Chen}, \bibinfo{person}{Chufeng Shi}, \bibinfo{person}{Ziwei Wu}, \bibinfo{person}{Jinfeng Yi}, {and} \bibinfo{person}{Xiangnan He}.} \bibinfo{year}{2020}\natexlab{}.
\newblock \showarticletitle{Model-Agnostic Counterfactual Reasoning for Eliminating Popularity Bias in Recommender System}.
\newblock \bibinfo{journal}{\emph{KDD}} (\bibinfo{year}{2020}).
\newblock


\bibitem[Wu et~al\mbox{.}(2022)]%
        {wu2022graph}
\bibfield{author}{\bibinfo{person}{Jiancan Wu}, \bibinfo{person}{Xiangnan He}, \bibinfo{person}{Xiang Wang}, \bibinfo{person}{Qifan Wang}, \bibinfo{person}{Weijian Chen}, \bibinfo{person}{Jianxun Lian}, {and} \bibinfo{person}{Xing Xie}.} \bibinfo{year}{2022}\natexlab{}.
\newblock \showarticletitle{Graph convolution machine for context-aware recommender system}.
\newblock \bibinfo{journal}{\emph{FCS}} (\bibinfo{year}{2022}).
\newblock


\bibitem[Wu et~al\mbox{.}(2020)]%
        {Wu2020SelfsupervisedGL}
\bibfield{author}{\bibinfo{person}{Jiancan Wu}, \bibinfo{person}{Xiang Wang}, \bibinfo{person}{Fuli Feng}, \bibinfo{person}{Xiangnan He}, \bibinfo{person}{Liang Chen}, \bibinfo{person}{Jianxun Lian}, {and} \bibinfo{person}{Xing Xie}.} \bibinfo{year}{2020}\natexlab{}.
\newblock \showarticletitle{Self-supervised Graph Learning for Recommendation}.
\newblock \bibinfo{journal}{\emph{SIGIR}} (\bibinfo{year}{2020}).
\newblock


\bibitem[Wu et~al\mbox{.}(2021)]%
        {fairgo}
\bibfield{author}{\bibinfo{person}{Le Wu}, \bibinfo{person}{Lei Chen}, \bibinfo{person}{Pengyang Shao}, \bibinfo{person}{Richang Hong}, \bibinfo{person}{Xiting Wang}, {and} \bibinfo{person}{Meng Wang}.} \bibinfo{year}{2021}\natexlab{}.
\newblock \showarticletitle{Learning fair representations for recommendation: A graph-based perspective}.
\newblock \bibinfo{journal}{\emph{WWW}} (\bibinfo{year}{2021}).
\newblock


\bibitem[Wu et~al\mbox{.}(2023)]%
        {citationsurveylekey}
\bibfield{author}{\bibinfo{person}{Le Wu}, \bibinfo{person}{Xiangnan He}, \bibinfo{person}{Xiang Wang}, \bibinfo{person}{Kun Zhang}, {and} \bibinfo{person}{Meng Wang}.} \bibinfo{year}{2023}\natexlab{}.
\newblock \showarticletitle{A Survey on Accuracy-Oriented Neural Recommendation: From Collaborative Filtering to Information-Rich Recommendation}.
\newblock \bibinfo{journal}{\emph{TKDE}} (\bibinfo{year}{2023}).
\newblock


\bibitem[Yang et~al\mbox{.}(2023b)]%
        {yang2023hyperbolic}
\bibfield{author}{\bibinfo{person}{Yonghui Yang}, \bibinfo{person}{Le Wu}, \bibinfo{person}{Kun Zhang}, \bibinfo{person}{Richang Hong}, \bibinfo{person}{Hailin Zhou}, \bibinfo{person}{Zhiqiang Zhang}, \bibinfo{person}{Jun Zhou}, {and} \bibinfo{person}{Meng Wang}.} \bibinfo{year}{2023}\natexlab{b}.
\newblock \showarticletitle{Hyperbolic Graph Learning for Social Recommendation}.
\newblock \bibinfo{journal}{\emph{TKDE}} (\bibinfo{year}{2023}).
\newblock


\bibitem[Yang et~al\mbox{.}(2023a)]%
        {yang2023generative}
\bibfield{author}{\bibinfo{person}{Yonghui Yang}, \bibinfo{person}{Zhengwei Wu}, \bibinfo{person}{Le Wu}, \bibinfo{person}{Kun Zhang}, \bibinfo{person}{Richang Hong}, \bibinfo{person}{Zhiqiang Zhang}, \bibinfo{person}{Jun Zhou}, {and} \bibinfo{person}{Meng Wang}.} \bibinfo{year}{2023}\natexlab{a}.
\newblock \showarticletitle{Generative-Contrastive Graph Learning for Recommendation}.
\newblock \bibinfo{journal}{\emph{SIGIR}} (\bibinfo{year}{2023}).
\newblock


\bibitem[Yu et~al\mbox{.}(2023)]%
        {Yao2020SelfsupervisedLF}
\bibfield{author}{\bibinfo{person}{Junliang Yu}, \bibinfo{person}{Hongzhi Yin}, \bibinfo{person}{Xin Xia}, \bibinfo{person}{Tong Chen}, \bibinfo{person}{Jundong Li}, {and} \bibinfo{person}{Zi Huang}.} \bibinfo{year}{2023}\natexlab{}.
\newblock \showarticletitle{Self-supervised learning for recommender systems: A survey}.
\newblock \bibinfo{journal}{\emph{TKDE}} (\bibinfo{year}{2023}).
\newblock


\bibitem[Yu et~al\mbox{.}(2022)]%
        {Yu2021AreGA}
\bibfield{author}{\bibinfo{person}{Junliang Yu}, \bibinfo{person}{Hongzhi Yin}, \bibinfo{person}{Xin Xia}, \bibinfo{person}{Tong Chen}, \bibinfo{person}{Li zhen Cui}, {and} \bibinfo{person}{Quoc Viet~Hung Nguyen}.} \bibinfo{year}{2022}\natexlab{}.
\newblock \showarticletitle{Are Graph Augmentations Necessary?: Simple Graph Contrastive Learning for Recommendation}.
\newblock \bibinfo{journal}{\emph{SIGIR}} (\bibinfo{year}{2022}).
\newblock


\bibitem[Zhang et~al\mbox{.}(2023b)]%
        {zhang2023invariant}
\bibfield{author}{\bibinfo{person}{An Zhang}, \bibinfo{person}{Jingnan Zheng}, \bibinfo{person}{Xiang Wang}, \bibinfo{person}{Yancheng Yuan}, {and} \bibinfo{person}{Tat-Seng Chua}.} \bibinfo{year}{2023}\natexlab{b}.
\newblock \showarticletitle{Invariant Collaborative Filtering to Popularity Distribution Shift}.
\newblock \bibinfo{journal}{\emph{WWW}} (\bibinfo{year}{2023}).
\newblock


\bibitem[Zhang et~al\mbox{.}(2023a)]%
        {zhang2023rethinking}
\bibfield{author}{\bibinfo{person}{Daoan Zhang}, \bibinfo{person}{Chenming Li}, \bibinfo{person}{Haoquan Li}, \bibinfo{person}{Wenjian Huang}, \bibinfo{person}{Lingyun Huang}, {and} \bibinfo{person}{Jianguo Zhang}.} \bibinfo{year}{2023}\natexlab{a}.
\newblock \showarticletitle{Rethinking alignment and uniformity in unsupervised image semantic segmentation}.
\newblock \bibinfo{journal}{\emph{AAAI}} (\bibinfo{year}{2023}).
\newblock


\bibitem[Zhang et~al\mbox{.}(2021)]%
        {zhang2021causal}
\bibfield{author}{\bibinfo{person}{Yang Zhang}, \bibinfo{person}{Fuli Feng}, \bibinfo{person}{Xiangnan He}, \bibinfo{person}{Tianxin Wei}, \bibinfo{person}{Chonggang Song}, \bibinfo{person}{Guohui Ling}, {and} \bibinfo{person}{Yongdong Zhang}.} \bibinfo{year}{2021}\natexlab{}.
\newblock \showarticletitle{Causal intervention for leveraging popularity bias in recommendation}.
\newblock \bibinfo{journal}{\emph{SIGIR}} (\bibinfo{year}{2021}).
\newblock


\bibitem[Zhao et~al\mbox{.}(2023b)]%
        {zhao2023fair}
\bibfield{author}{\bibinfo{person}{Chen Zhao}, \bibinfo{person}{Le Wu}, \bibinfo{person}{Pengyang Shao}, \bibinfo{person}{Kun Zhang}, \bibinfo{person}{Richang Hong}, {and} \bibinfo{person}{Meng Wang}.} \bibinfo{year}{2023}\natexlab{b}.
\newblock \showarticletitle{Fair Representation Learning for Recommendation: A Mutual Information Perspective}.
\newblock \bibinfo{journal}{\emph{AAAI}} (\bibinfo{year}{2023}).
\newblock


\bibitem[Zhao et~al\mbox{.}(2023a)]%
        {Popularity-awareDRO}
\bibfield{author}{\bibinfo{person}{Jujia Zhao}, \bibinfo{person}{Wenjie Wang}, \bibinfo{person}{Xinyu Lin}, \bibinfo{person}{Leigang Qu}, \bibinfo{person}{Jizhi Zhang}, {and} \bibinfo{person}{Tat-Seng Chua}.} \bibinfo{year}{2023}\natexlab{a}.
\newblock \showarticletitle{Popularity-aware Distributionally Robust Optimization for Recommendation System}.
\newblock \bibinfo{journal}{\emph{CIKM}} (\bibinfo{year}{2023}).
\newblock


\bibitem[Zhao et~al\mbox{.}(2022b)]%
        {Zhao2022InvestigatingAP}
\bibfield{author}{\bibinfo{person}{Minghao Zhao}, \bibinfo{person}{Le Wu}, \bibinfo{person}{Yile Liang}, \bibinfo{person}{Lei Chen}, \bibinfo{person}{Jian Zhang}, \bibinfo{person}{Qilin Deng}, \bibinfo{person}{Kai Wang}, \bibinfo{person}{Xudong Shen}, \bibinfo{person}{Tangjie Lv}, {and} \bibinfo{person}{Runze Wu}.} \bibinfo{year}{2022}\natexlab{b}.
\newblock \showarticletitle{Investigating Accuracy-Novelty Performance for Graph-based Collaborative Filtering}.
\newblock \bibinfo{journal}{\emph{SIGIR}} (\bibinfo{year}{2022}).
\newblock


\bibitem[Zhao et~al\mbox{.}(2022a)]%
        {evil}
\bibfield{author}{\bibinfo{person}{Zihao Zhao}, \bibinfo{person}{Jiawei Chen}, \bibinfo{person}{Sheng Zhou}, \bibinfo{person}{Xiangnan He}, \bibinfo{person}{Xuezhi Cao}, \bibinfo{person}{Fuzheng Zhang}, {and} \bibinfo{person}{Wei Wu}.} \bibinfo{year}{2022}\natexlab{a}.
\newblock \showarticletitle{Popularity bias is not always evil: Disentangling benign and harmful bias for recommendation}.
\newblock \bibinfo{journal}{\emph{TKDE}} (\bibinfo{year}{2022}).
\newblock


\bibitem[Zheng et~al\mbox{.}(2021)]%
        {zheng2021disentangling}
\bibfield{author}{\bibinfo{person}{Yu Zheng}, \bibinfo{person}{Chen Gao}, \bibinfo{person}{Xiang Li}, \bibinfo{person}{Xiangnan He}, \bibinfo{person}{Yong Li}, {and} \bibinfo{person}{Depeng Jin}.} \bibinfo{year}{2021}\natexlab{}.
\newblock \showarticletitle{Disentangling user interest and conformity for recommendation with causal embedding}.
\newblock \bibinfo{journal}{\emph{WWW}} (\bibinfo{year}{2021}).
\newblock


\bibitem[Zhu et~al\mbox{.}(2019)]%
        {zhu2019aligning}
\bibfield{author}{\bibinfo{person}{Yongchun Zhu}, \bibinfo{person}{Fuzhen Zhuang}, {and} \bibinfo{person}{Deqing Wang}.} \bibinfo{year}{2019}\natexlab{}.
\newblock \showarticletitle{Aligning domain-specific distribution and classifier for cross-domain classification from multiple sources}.
\newblock \bibinfo{journal}{\emph{AAAI}} (\bibinfo{year}{2019}).
\newblock


\bibitem[Zhu et~al\mbox{.}(2021a)]%
        {Zhu2021PopularityOpportunityBI}
\bibfield{author}{\bibinfo{person}{Ziwei Zhu}, \bibinfo{person}{Yun He}, \bibinfo{person}{Xing Zhao}, \bibinfo{person}{Yin Zhang}, \bibinfo{person}{Jianling Wang}, {and} \bibinfo{person}{James Caverlee}.} \bibinfo{year}{2021}\natexlab{a}.
\newblock \showarticletitle{Popularity-Opportunity Bias in Collaborative Filtering}.
\newblock \bibinfo{journal}{\emph{WSDM}} (\bibinfo{year}{2021}).
\newblock


\bibitem[Zhu et~al\mbox{.}(2021b)]%
        {zhu2021progressive}
\bibfield{author}{\bibinfo{person}{Zhen Zhu}, \bibinfo{person}{Tengteng Huang}, \bibinfo{person}{Mengde Xu}, \bibinfo{person}{Baoguang Shi}, \bibinfo{person}{Wenqing Cheng}, {and} \bibinfo{person}{Xiang Bai}.} \bibinfo{year}{2021}\natexlab{b}.
\newblock \showarticletitle{Progressive and aligned pose attention transfer for person image generation}.
\newblock \bibinfo{journal}{\emph{TPAMI}} (\bibinfo{year}{2021}).
\newblock


\end{thebibliography}
\end{document}